# Seismic noise suppression: array stations, waveform cross-correlation, and noise stochastization

Ivan O. Kitov

**Abstract**

Seismic noise with an amplitude higher than that of the sought signal is a challenge for detection. Several techniques have been developed to suppress the ambient noise and to reduce the detection threshold in order to find signals with the lowest possible amplitudes produced by events with the magnitudes significant for scientific research and technical applications. The first technique used to find signals is filtering. It is most efficient when there is a frequency range where the signal is above noise. Seismic arrays were introduced in the late 1950s as a method for improving underground test monitoring, potentially reducing detection thresholds by fivefold or more by exploiting destructive interference effects of a quasi-random noise. The beamforming method is the backbone of data processing at the International Data Centre (IDC) with more than 30 array stations of the International Monitoring System (IMS) installed around the globe. The matched filter method allows for the suppression of noise incoherent to the sought signal. It employs waveform cross-correlation (WCC) with templates based on actual and simulated seismic signals to improve the signal-to-noise ratio estimates for similar signals. The performance of this method is significantly enhanced when it is applied to a seismic array. A novel technique combined with WCC, is the noise stochastization or the addition of scaled random noise to the actual data before calculating the cross-correlation coefficient. The stochastic component can easily be generated by a computer program. Alternatively, a regular signal propagating at an angle of around 90° to the plane of the sought signal can play a role of stochastic component at array stations. We demonstrate the separate and joint effects of these noise reduction techniques on the WCC performance, when applied to filtered data from selected IMS arrays and various waveform templates of historical events available at the IDC.

## Introduction

The main objective of a seismological agency is to recover seismic activity within a defined area, to the extent allowed by the available data. Finding seismic events of various natures and estimating their parameters are two routine operational tasks. Signal detection is the first step in event finding and, since the advent of digital seismology in the 1970s, has heavily relied on automated processing. Valid seismic signals are searched for among other signals, which, being also fully physical and having their own sources, are not considered as valid. These "false" signals create the ambient seismic noise and can include many waveforms similar to the signals accepted as valid. They are considered false because they are embedded in seismic noise with a higher amplitude and do not have high enough signal-to-noise ratio or other quality characteristics required for a valid signal. The limit case often encountered in practice is when the noise consists of coherent signals and there is no algorithm to separate overlapping valid and false cases. This case is especially important for array seismic stations where the reduction in the detection threshold is based on the destructive interference of noise and constructive interference of signals at individual sensors distributed over the earth surface.

All in all, seismic noise realizations with the same stationary power can vary from a fully coherent to the sought valid signal (*e.g.,* seismic noise at station MJAR (Japan) right after the Tohoku earthquake) to an almost random case with a very low-amplitude component coherent to any valid regular signal generated by the sources of interest. There is no physical meaning of valid/false signal separation and the rational comes from event characterization and statistical properties of valid event hypotheses. The latter can be obtained from historical observations in a given area and the areas with similar geology and seismicity. Limited human resources are also a parameter that defines detection thresholds and the criteria for valid events. Any advances in the characterization of the detection quality and the reduction of the detection threshold enhance the capabilities of seismic observation, resulting in the improvements in the catalogue/bulletin completeness, consistency and accuracy.

The International Monitoring System (IMS) of the Comprehensive Nuclear-Test-Ban Treaty Organization (CTBTO) includes more than 30 seismic arrays of various sizes (aperture from ~200 m to 42 km) and distributed over Eurasia (26), North America (5), Australia (2) and one station TORD is in Africa [Coyne at al., 2012]. These arrays are the most efficient tool for detecting the low-magnitude events that are relevant to the CTBTO mandate [CTBT, 1996], as they allow reducing the detection threshold significantly. The IMS transfers raw data segments to the International Data Centre (IDC). There, they are first processed in automatic regime. An interactive review of the automatic bulletin (Standard Event List 3, SEL3) follows strict rules to guarantee the statistical significance of seismic events in the final IDC internal product - Reviewed Event Bulletin (REB). The REB quality rules are based on quantitative Event Definition Criteria (EDC) [Coyne *et al*., 2012] scrupulously estimated by numerous international experts during the CTBTO preparation stage [Ringdal, 1994]. The efficiency of the CTBT monitoring regime critically depends on the detection threshold as all seismic (infrasound, hydroacoustic, and RN) events in the IMS data have to be found, characterized, and made available for the States Signatory. The IMS has the best global seismic network with high quality and real time data to meet these requirements.

The quality of routine processing at the IDC is thoroughly tested and improved. One of many testing procedures involves in-depth investigations into specific events, regions, and time periods. There is a Spot Check Tool (SCT) designed to provide advanced analysis of seismic data [Kitov, Dricker, 2025]. The SCT methods and algorithms were selected and implemented not only to test the quality of routine products such as the REB and SEL3, but also to extend the list of valid events in both bulletins. The set of SCT tools and techniques has been progressively extended to incorporate the most recent developments in seismic observations and data processing. One of the directions of the SCT enhancement is the usage of the matched filter technique [Turin, 1960] for detection and the local association (LA) approach for creation of reliable event hypotheses matching the EDC for automatic and interactive products [Bobrov *et al*., 2014]. These advanced methods allow to test standard processing algorithms and improve their performance in a positive feed-back loop.

The introduction of waveform cross correlation (WCC) into seismological practice started in the 1990s [Israelsson, 1990; Joswig, 1990; Joswig, Schulte-Theis, 1993] and regained its momentum in the 2000s [Schaff, Richards, 2004; Gibbons, Ringdal, 2006, 2008; Waldhauser, Schaff, 2008; Schaff, 2009]. The WCC has been expanding as a detection method ever since [Gibbons *et al*., 2011, 2017; Adushkin *et al*., 2015, 2017; 2025], and now it is thoroughly used by various institutions, organizations, and seismological agencies [Schaff, Waldhauser, 2010; Schaff, Richards, 2014; Bobrov *et al*., 2014; 2016ab; 2017; Herrmann *et al*., 2019; Mu *et al*., 2019; Beaucé, *et al.,* 2023; Mesimeri *et al*., 2024; Kitov, Rozhkov, 2024; Kitov *et al*., 2025].

A number of REB events together with their corresponding signals were selected according to their quality and their cross-correlation characteristics with all other REB events within 3°. The selected events (location, origin time, $m_b$ and Ms magnitudes) are known as master events (ME). The signals associated with these events are used as waveform templates in the matched filter. A few exercises with interactive review conducted at the IDC with the WCC have shown excellent results. The usage of the cross-correlation standard event list (XSEL) instead of its standard analog - automatic SEL3 bulletin, allowed to find from 50% to 100% more REB events in addition to the official REB [Bobrov *et al*., 2014; Bobrov *et al*., 2016a]. The XSEL is machine learning friendly as its events are repeating known REB events effectively used for training [Bobrov *et al*., 2017] and synthetic seismograms are almost as effective in detection as real ones [Bobrov *et al.,* 2016b].

In this paper, we present selected results of WCC processing of the IMS data at array stations from earthquakes and different noise conditions in order to demonstrate the advantages and variations in the performance of WCC-based detection methods. The introduction of these methods into routine processing starts with an extensive testing in the environment for the in-depth analysis of various cases - from the detection of ultra-weak signals from low-magnitude seismic events [Adushkin *et al*., 2015, 2017, 2025; Kitov, Sanina, 2022, 2025], to the detection of specific signals in high-amplitude noise coherent to the sought signal [Kitov *et al*., 2026].

**Data, new methods, results**

*General detection features*

The previous versions of WCC processing was limited to the direct calculation of cross-correlation coefficients (*CC*) at three-component (3-C) and array stations of the IMS [Bobrov *et al*., 2014, 2017]. The *CC*-traces are used to calculate corresponding SNRcc as related to the matched filter detector. To define the optimal detection thresholds, band-pass filters and cross-correlation window widths (CCWW) for any given ME at all associated stations, a wider range of detection statistics of valid/false arrivals was collected and analyzed. Valid are those detections that associated with the event hypotheses in the XSEL. False detections are, in turn, those that were not associated due to various reasons. The effects of a comb of band-pass filters applied to raw seismic data, and the CCWW, were also collected for gather statistics and for further tuning of the final processing version.

The IMS data and the REB are two sources of raw data and metadata needed for the detailed analysis of new algorithms and techniques aimed at the overall improvement of the WCC-related detection. The previous WCC version has been enhanced by new features to reduce detection threshold and reliability of detections/event hypotheses. For beamforming technique, array stations have a beam loss problem related to the difference between the theoretical and empirical time delays between arrival times of real waves propagating along them. The WCC processing does not meet this problems as it implies the close proximity of the ME and the sought events making all arrivals obtained by WCC at all sensors to be fully synchronized in theory. Nevertheless, data discretization, the difference between MEs and sought events positions and seismic noise affecting low-amplitude sought signals all can shift the observed arrival times from their expected positions. As a remedy, a set random shifts (shakes) of all arrival times at individual sensors by one-three counts is introduced to individual CC-traces. If the CC-traces are shifted relative to their real positions, such a procedure can have some realization of random shifts which create a set much closer to the optimal set. In this case, the sum of individual CC-traces averaged over all working channels may increase significantly near the expected peak and the corresponding SNRcc value will increase. In some cases, new detections can be found if the SNRcc grows above the threshold. To avoid random spikes in the averaged *CC* time series and in the corresponding SNRcc, the minimum allowed gain in the SNRcc is defined. This "shake" method has been tested for a few years and provides a significant increase in the number of detections and their SNRcc values. This improves the XSEL completeness and consistency.

*Noise suppression by seismic arrays*

The array stations of the IMS provide a significant improvement in detection of weak signals relative to the collocated 3-C stations due to effective noise suppression. Theoretically, the constructive interference of fully coherent signals at individual sensors without energy loss (beam loss can be high in reality) [Schweitzer *et al*., 2012] and the destructive interference of the random noise result in the signal detection average gain proportional to the square root of the number of channels. Unfortunately, the microseismic noise always contain components which are partially coherent to the template and sought signals. The destructive noise interference is not working when the coherent noise is high. The ultimate case of such a problem is the signals from aftershocks of the largest earthquakes. The mainshock with duration of tens of seconds and the length of hundreds of kilometers rise the noise level by orders of magnitude. No aftershocks can be found in the first several minutes following the catastrophic earthquakes. The IMS and IDC are effectively blinded for several minutes and the blind spot can be thousand of kilometers.

The noise fully coherent to the sought and template signals is very difficult to suppress by simple beamforming [Schweitzer *et al*., 2012] and basic cross-correlation methods [Adushkin *et al.*, 2025a]. There are some opportunities to improve the performance of WCC processing by adding a stochastic noise component to the coherent noise. There are two general approaches for the array stations.

*Changing incident azimuths of regular plane waves*

For an array station, the coherence of a template and the sought signal depends on angle between their respective propagation directions and apparent velocities. The coherence decays quickly when the plane waves arrive at different angles and/or have different scalar slownesses. Two signals travelling in orthogonal directions along an array and having different slownesses are practically incoherent. They represent an almost random noise to each other when the WCC method is applied [Kitov, Sanina, 2025b]. For example, the SNRcc can be improved if to add a signal from the Tohoku earthquake to the signals from an earthquake close to the Tohoku aftershock area as observed at IMS stations.

Instead of searching for an actual signal arriving from the direction orthogonal to the sought signal it is possible to turn any multichannel signal at a given array by changing the relative arrival times at individual sensors to the theoretical set corresponding to the needed direction and apparent velocity. For example, one can turn the Tohoku wave plane to a direction orthogonal to that of the wave from another close to the Tohoku event. The Tohoku signal can operate as coherent noise or incoherent stochastic noise depending on the incident angle. The amplitude of the turned noise signal can be scaled by using the multiplication factor *SeisN*, which can be any value, including *SeisN*=0, the "no-noise" case. Figure 1 shows the SNRcc peak value for the signal from the March 16, 2022 earthquake at IMS station CMAR as a function of the incident angle of the turning Tohoku signal used as stochastic noise (see Table 1 for the events parameters).

Table. 1. Parameters determined by the IDC for two earthquakes in Figure 1. The third earthquake occurred 126 s before the second one and was very close to it. Sought event is highlighted bold.

| IDC orid | Date | Origin time | Lat, deg | Lon, deg | Depth, km | $m_b$ | ndef | Smaj, km | Smin, km |
|---|---|---|---|---|---|---|---|---|---|
| 7326479 | 11.03.2011 | 05:46:19.32 | 38.435 | 142.520 | 0 | 5.51 | 122 | 8.0 | 6.9 |
| **21963352** | **16.03.2022** | **14:36:32.94** | **37.709** | **141.548** | **59.4** | **6.05** | **103** | **7.2** | **5.9** |
| 21962192 | 16.03.2022 | 14:34:26.62 | 37.662 | 141.667 | 53.4 | 5.22 | 66 | 10.5 | 7.8 |

The station-event azimuth for the 16.03.2022 event is 54° and the bottom of the SNRcc trough in Figure 1a is around 51° with the angle step of 6°. The SNRcc is much higher for all other angles than for the case of "no-noise" (red line), *i.e.* the Tohoku noise amplitude factor *SeisN*=0. The amplitude factor allows to find the best case for the highest SNRcc: *SeisN*=0.05 and the incidence angle of ~180°. There is a wide shelf of very close SNRcc values starting at approximately 150°, i.e. approximately perpendicular to the direction from the station to the 16.03.2022 event. The distribution of 20 sensors for station CMAR is not even and its sensitivity may be azimuth dependent and the SNRcc peak is slightly shifted to the direction of maximum sensitivity. The *SeisN* values lower than 0.05 provide some SNRcc growth relative to the *SeisN*=0 case, but the suppression is not optimal and the coherent component in the real ambient noise still affects the matched filter performance. For *SeisN* values higher than 0.05, the effect of high-amplitude seismic noise added to the real data prevails and the WCC also underperforms relative to the peak value. We do not know in advance the level of the coherent noise component in the real noise observed on March 16, 2022. Therefore, we have to use the whole range near the peak value when processing with the added noise, which should be preferably a regular P-phase (only P-phases are used as templates) signal long enough to represents the noise with long-term-average (LTA) of 60 s. Any other seismic phase as well as microseismic noise can be added with an appropriate *SeisN* factor.

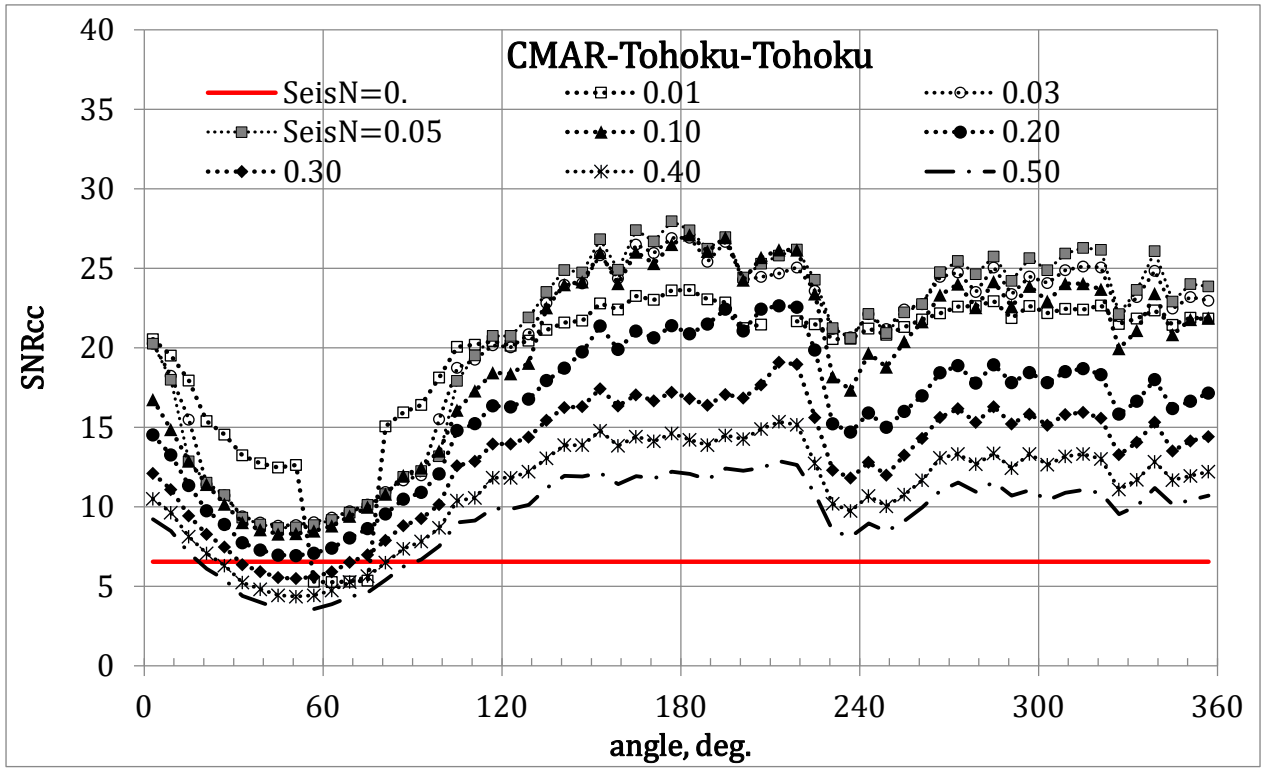

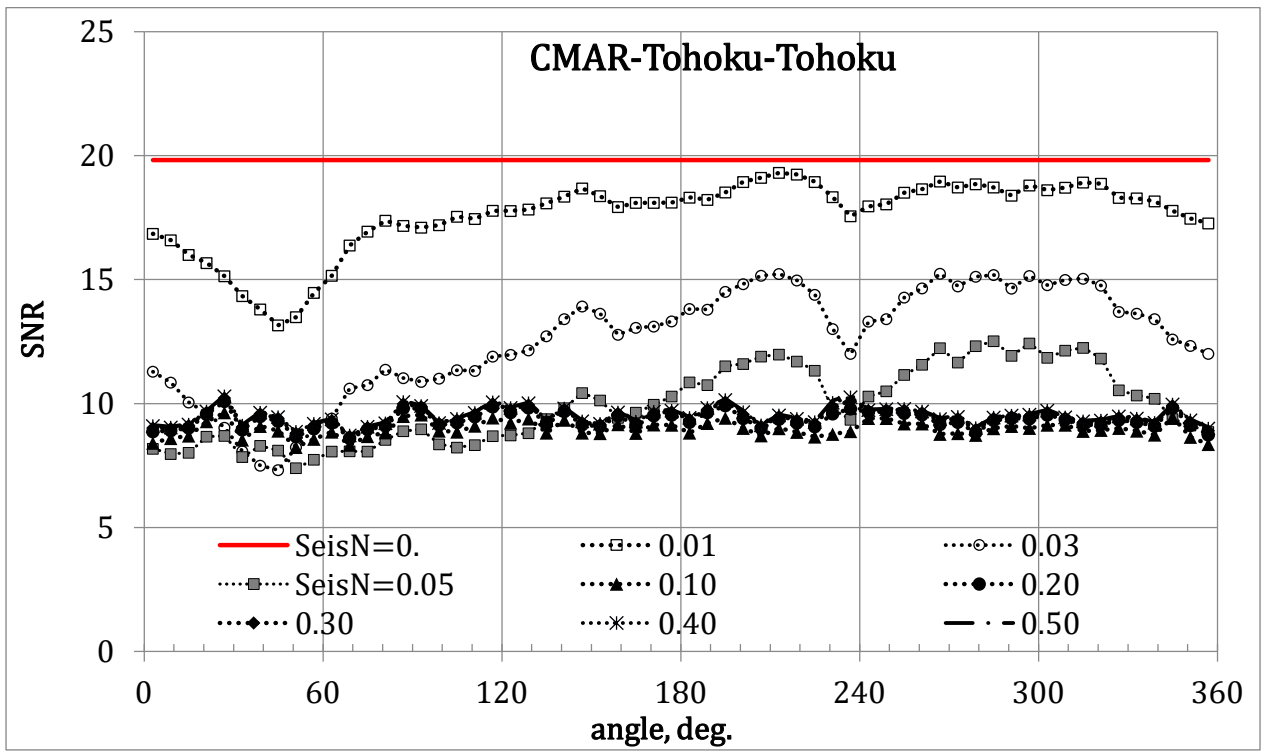

Figure 1. SNRcc (a) and SNR (b) as a function of Tohoku signal incidence angle for the 16.03.2022 event.

The SNR curve in Figure 1b demonstrates the overall decrease with the increase in *SeisN*. This is an illustration how the beamforming works in the Tohoku noise. There is the SNR dependence on the incidence angle as well with a deep trough around 54°.

The main outcome of this example is that the (reduced in amplitude) Tohoku signal can destroy the noise component coherent to the signal from the events in the Tohoku aftershock area at IMS stations. At the same time, the Tohoku earthquake (or any other catastrophic earthquake) was, could be, or will be if repeated, a real source of noise for many event masking their detection. The angle dependence of the WCC performance in these cases deserves further investigation and implementation in the in-depth analysis of IMS data.

*Noise suppression by digital stochastization*

Adding numerically generated stochastic noise of varying amplitude to the event recordings (also known as noise "whitening") before WCC processing is a more effective and flexible way to destroy coherent noise [Adushkin *et al*., 2025a]. With an appropriate level of the added stochastic noise amplitude define by multiplication factor *StochN,* the component of the ambient noise coherent to the sought and template signals approaches the stochastic noise properties as defined by the matched filter condition for the optimal detector based on the signal-to-noise ratio. We use factor *StochN* to vary the noise amplitude, which defines the amplitude of digital stochastic noise relative to the peak amplitude of the waveform in a given data segment. For the studied cases, the peak amplitude belongs to the sought signal from the earthquake. This is a convenient way to scale the stochastic noise level over the whole studied time interval. There are other possible scaling procedures. For example, to modulate the stochastic noise amplitude with the LTA or short-term-average (STA) level representing the noise and signal amplitudes. This allows to tune the *StochN* level to the level of coherent noise locally [Kitov *et al*., 2026b].

Figure 2 shows the case of the March 16, 2022 earthquake for station ASAR. The "*StochN*=0" curves present the original WCC processing without any change in the waveform. The SNR curve falls below the start level and reaches the detection threshold (SNR=4.0) for *StochN* near 1, *i.e.* when the stochastic noise reaches the peak signal level. The stochastic noise effectively suppresses the original signal in the beamforming procedure. The SNRcc starts to increase slowly with increasing *StochN* as the destructive interference of the coherent noise starts immediately. The peak of SNRcc of 55 relative to the start value of ~42 is observed at *StochN*=0.17. Then the SNRcc curve starts to fall, passing the start value at *StochN*~0.45, and slowly reaches the detection threshold at *StochN*~10. The latter value means that WCC can find signal in the stochastic noise 10 time higher in amplitude(!). The advantage of the matched filter before beamforming is additionally illustrated in Figure 3 for station AKASG.

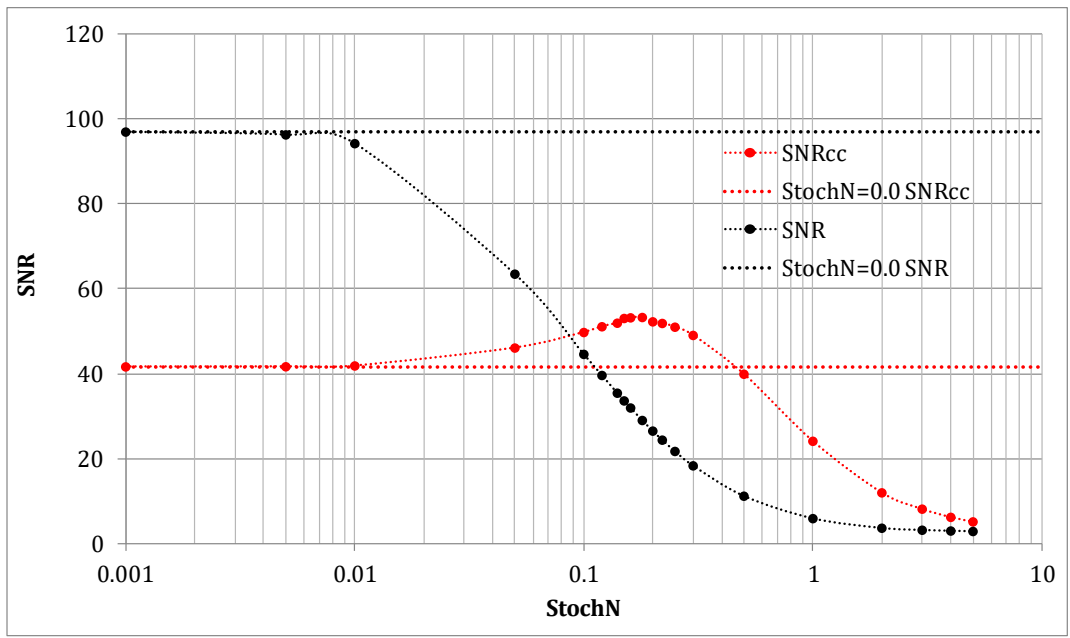

Figure 2. The peak SNRcc and SNR curves for the case with *SeisN*=0, angle=0°, and increasing *StochN* at station ASAR. The SNRcc curve peaks at 55 with *StochN*=0.17 and then fall below the "no-noise" level (41.6) at *StochN*~0.45.

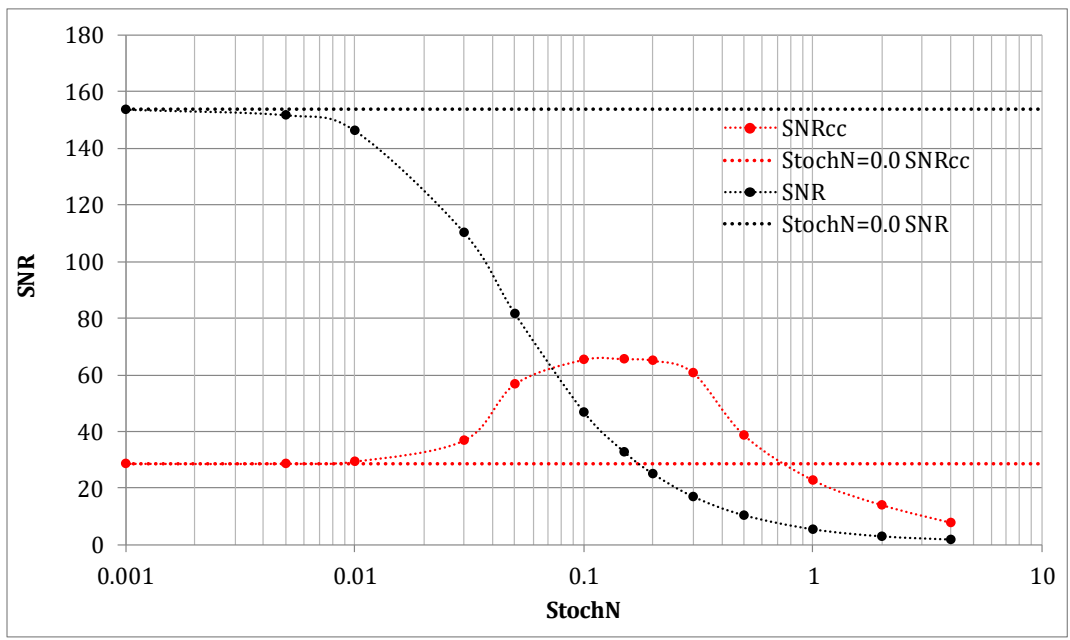

Figure 3. Same as in Figure 2 for AKASG.

## *Joint usage of digital stochastization and adding regular waves*

Regular seismic signals and stochastic noise both are able to destroy the noise component coherent to the template and sought signals, and thus improve the estimates of SNRcc. Their efficiency depends on the relative amplitude of the coherent noise component. At the same time, both methods used to destroy the coherent noise component severely affect the standard methods of signal detection and SNR falls when such noise is added. The energy detector is less sensitive to the suppression of the signal coherent noise and more to the noise amplitude itself. A regular seismic signal used as an instrument to increase the efficiency of a coherent noise destructive interference is sensitive to the direction of a plane wave of the used signal. To achieve the optimal efficiency, the plane has to be orthogonal to the propagation direction of coherent component in the ambient noise. Otherwise, it can increase the amplitude of the coherent noise. In real cases of high-amplitude noise, the sought signal is usually immersed in regular signals from the events far from the corresponding ME position. This is observed immediately after the great earthquakes. The high-amplitude noise represented by regular signals from the remote catastrophic earthquake is the noise to suppress as it may contain a component coherent to the sought signal and thus deteriorate the detection capabilities of beamforming and WCC. This is the case when the stochastic noise can and should be used to destroy the regular-phase transient noise. Therefore, we test the joint effect of the two above methods when the regular phase noise is not orthogonal to the sought signal. They can be orthogonal or even collinear. These cases should be also studied.

The SNRcc curve in Figure 1 shows a non-linear dependence on the angle between the sough and regular-phase-noise planes. The stochastic noise effect is also not linear as a function on noise relative amplitude as per Figures 2 and 3. Therefore, the interaction of two methods should be also nonlinear and the stochastic noise level needed to optimally suppress the effect of the regular-phase-noise can depend on the angle. This is important for the in-depth analysis as the most difficult cases are those that involve the detection of signals weak relative to the ambient noise amplitude whatever is the noise structure - random mixture of a large number of various signals from different sources or a transient regular phase from a remote earthquake. The capability to reduce the detection threshold is such cases are crucial for the CTBTO mission.

Figure 4 shows waveforms at IMS station ASAR obtained from the Toholu earthquake and the March 16, 2022 smaller earthquake (see Table 1). There was another earthquake approximately 2 minutes before the 16.03.2022 event very close to it. This smaller earthquake generated signals which represent the noise coherent to the signal from the sought event. So, we know that the noise before the sought signal is coherent to it but lower in amplitude. This is the noise to suppress. This noise can be changed in amplitude if we move the signal from noise signal from the earlier earthquake closer (40 s) to the sought signal (middle trace in Figure 4) and add it channel-by-channel to the original waveform. The sought earthquake is used as a ME and that fact provides the ultimate autocorrelation case which is the best for the estimation of the method efficiency as cross-correlation itself is impeccable.

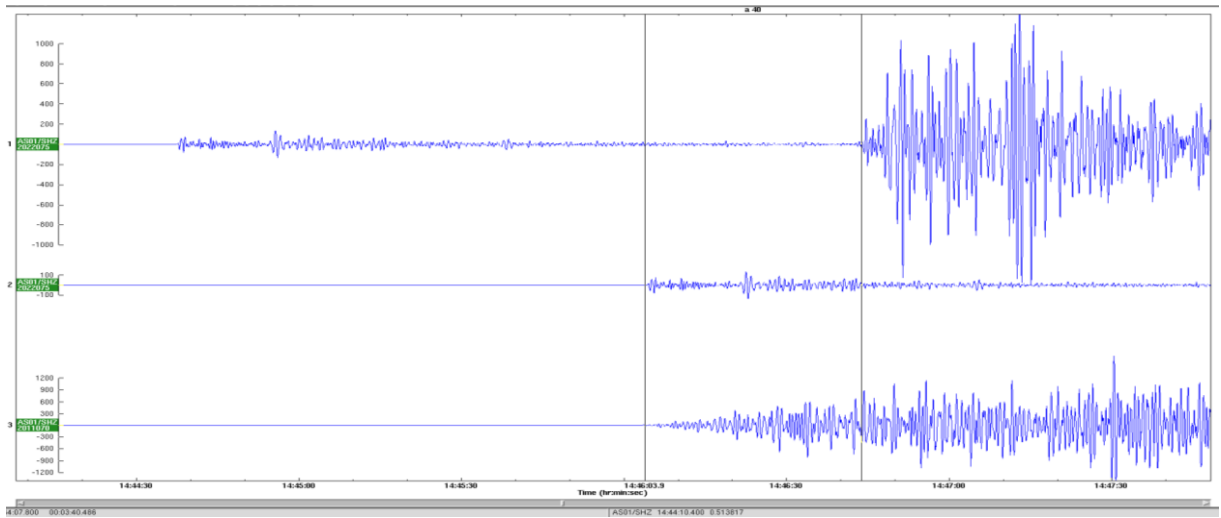

Figure 4. Signals from Tohoku (jdate=2011070) and two collocated earthquakes on 2022075 at ASAR.

In Figure 5, several cases of the SNRcc dependence on the noise-plane-wave (from Tohoku) angle relative to the sought signal are presented for various pairs of *SeisN* and *StochN* values. These cases simulate the positions, origin times, and magnitudes of the Tohoku earthquake relative to the sought event on March 16, 2022. The amplitude of the added natural noise is defined by *SeisN* relative to the real Tohoku peak amplitude value at station ASAR. The amplitude of stochastic noise, *StochN*, is scaled to the amplitude of the sought signal. For *StochN*=0.2 and 0.3, the case of *SeisN*=0.0005 is successful for the stochastic noise suppression effect and the SNRcc curves are above the "no-noise" case when *SeisN*=0.0 and *StochN*=0.0. When *SeisN*>0.001, the stochastic noise cannot improve the "no-noise" case.

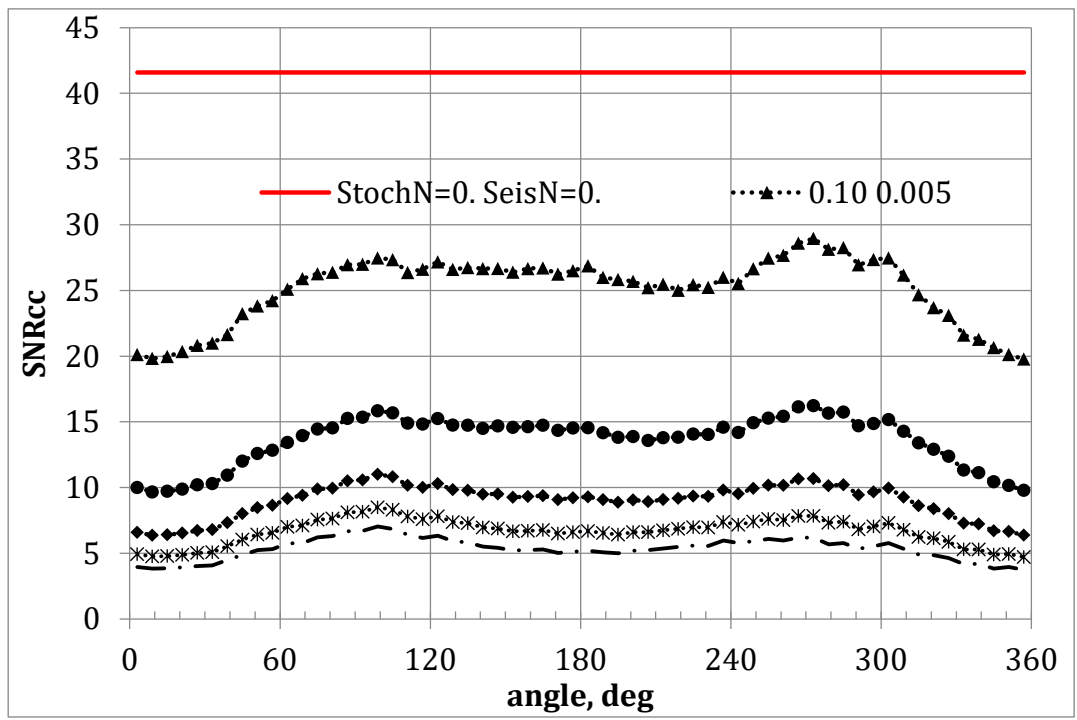

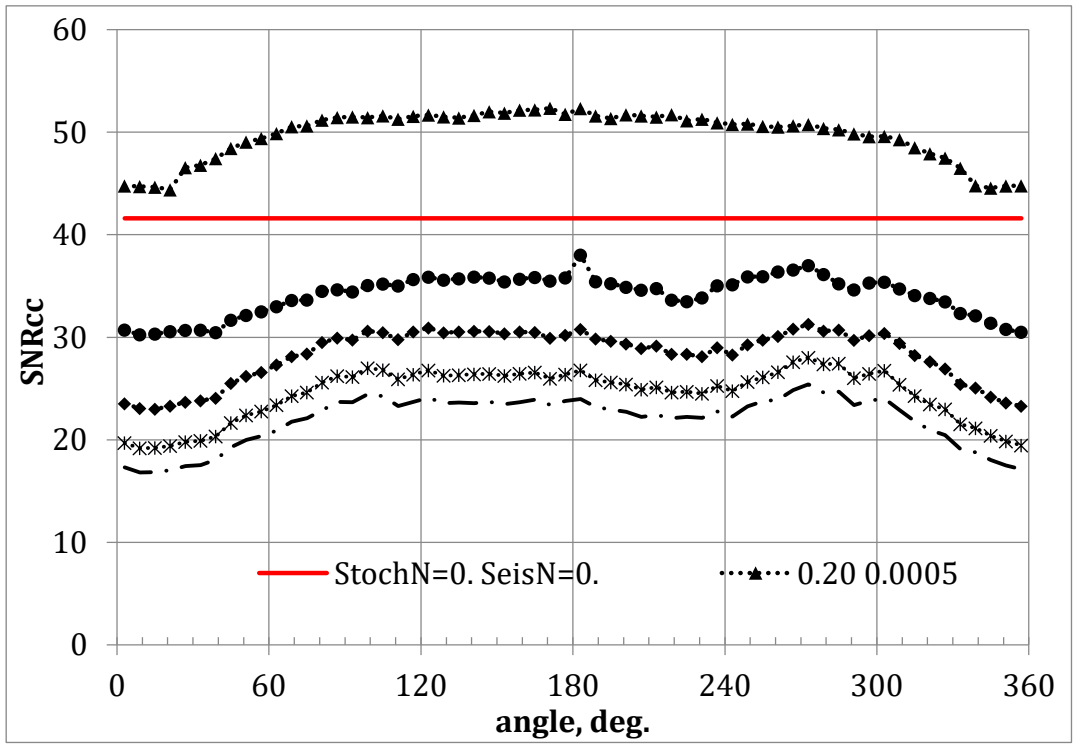

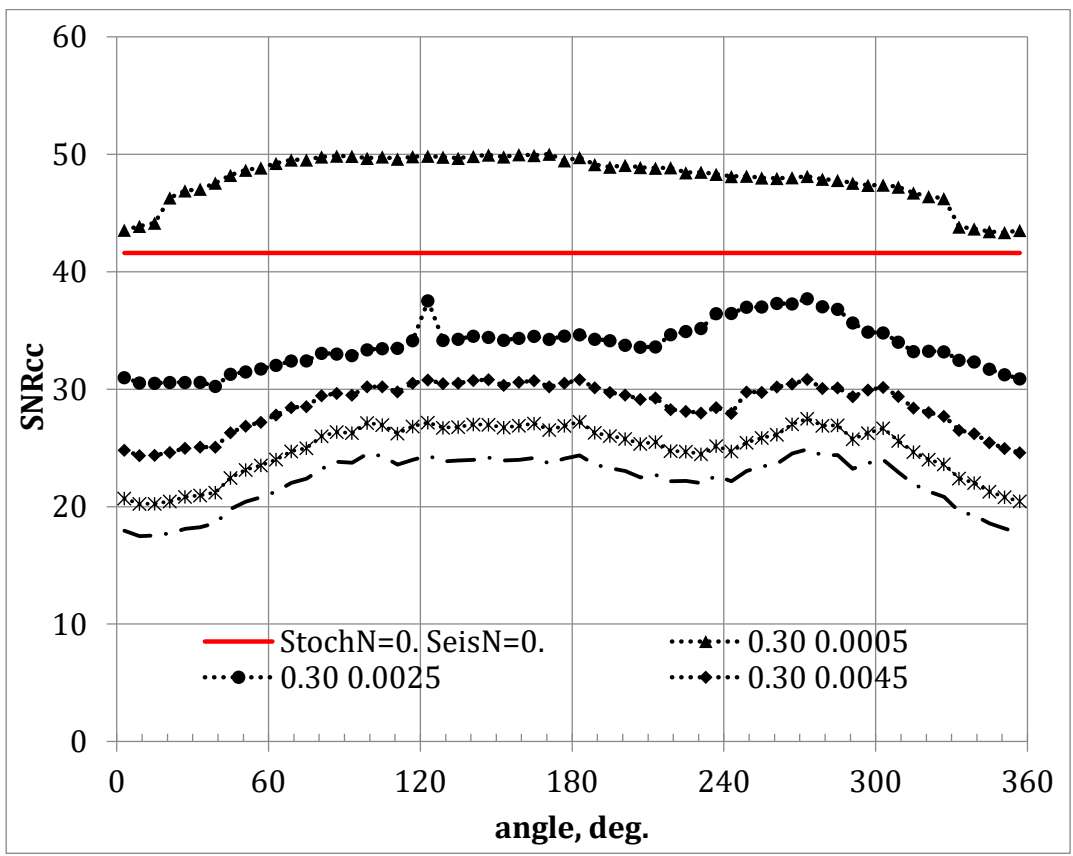

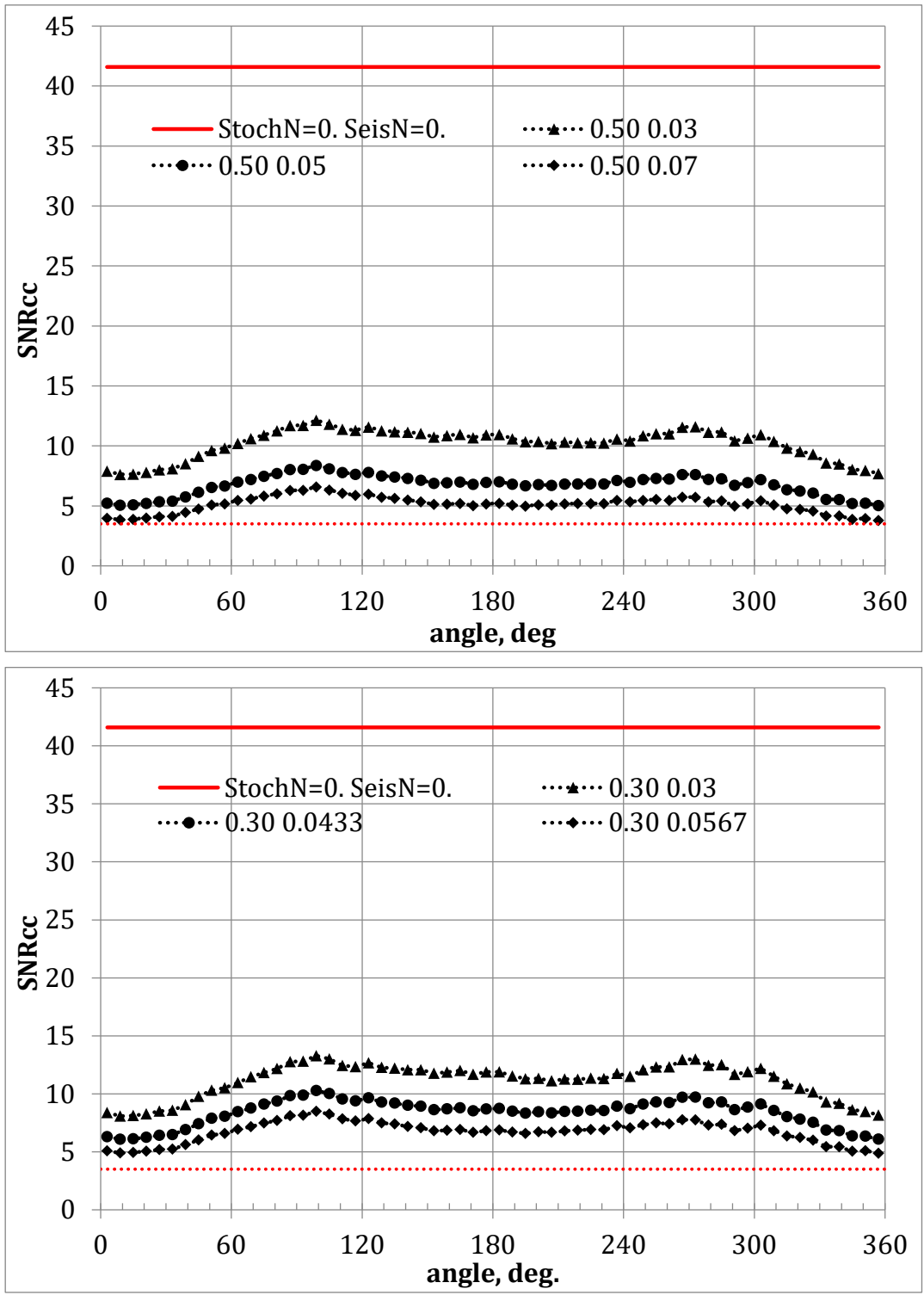

Figure 5. The SNRcc evolution with increasing *SeisN* and *StochN* depending on the Tohoku transient noise incidence angle. SNRcc detection threshold is of 3.5.

Figure 6 presents a case similar to that in Figure 5 for IMS array station AKASG. The relative amplitudes of the Tohoku mainshock and 16.03.22 earthquake are different from station ASAR. For *SeisN*=0.0005 and *StochN*=0.1, the increase in SNRcc relative to the "no-noise" case is dramatic - by a factor of 2. There is a deep trough at 55° where the Tohoku plane is collinear to the observed signals from March 16, 2022 event. With increasing *SeisN* and constant *StochN*=0.1, the gain related to the adding of the stochastic noise component decreases to the "no-noise" level and below for various angles. The SNRcc curves rise when a larger *StochN*=0.2 is used.

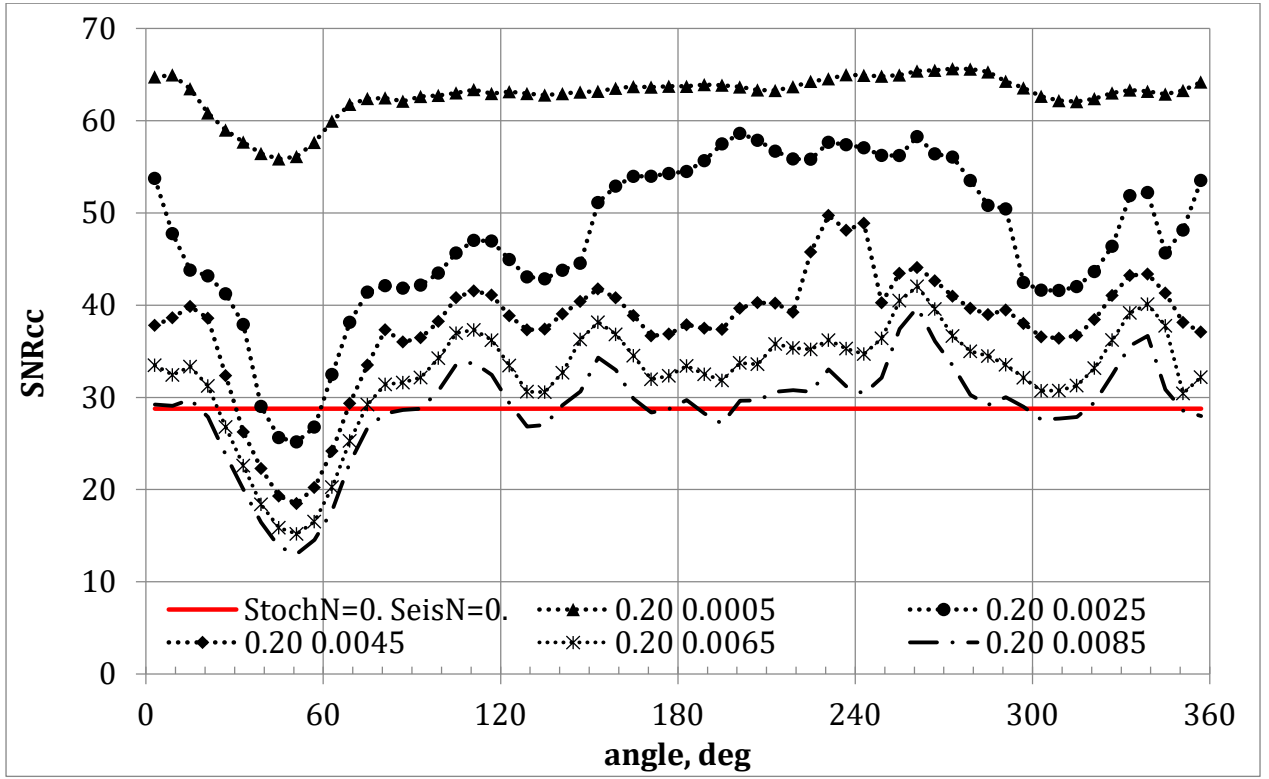

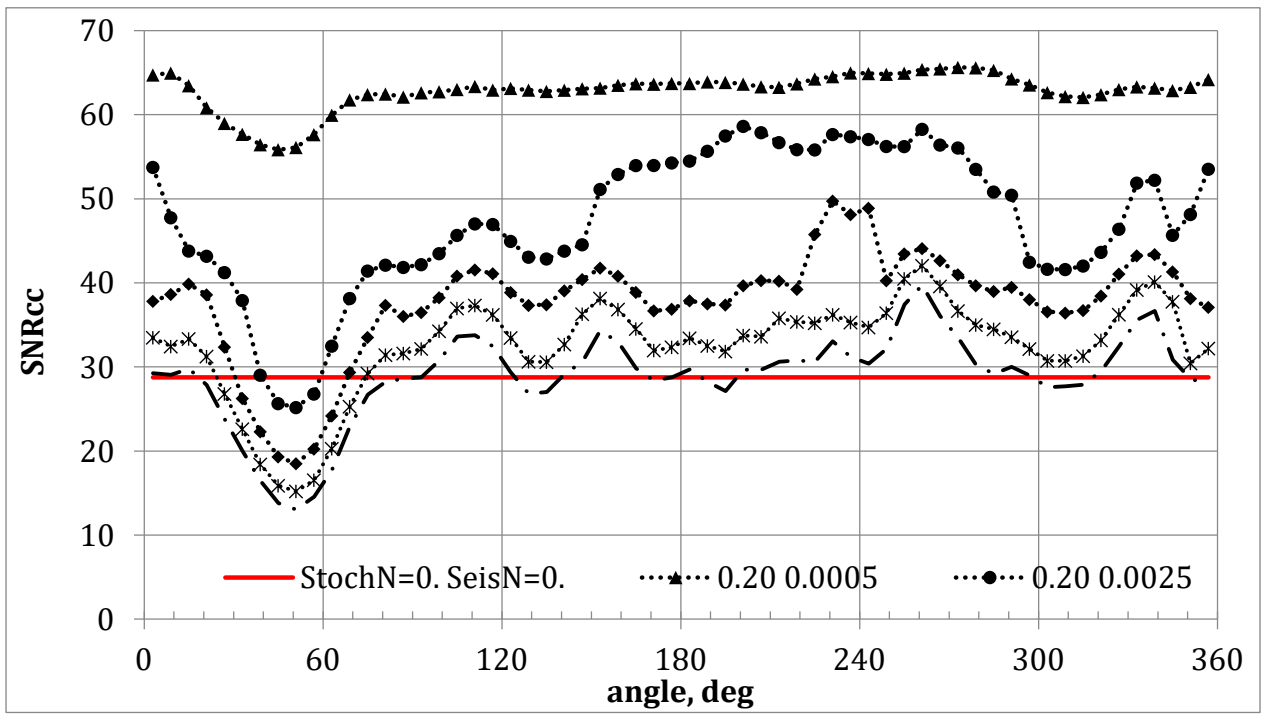

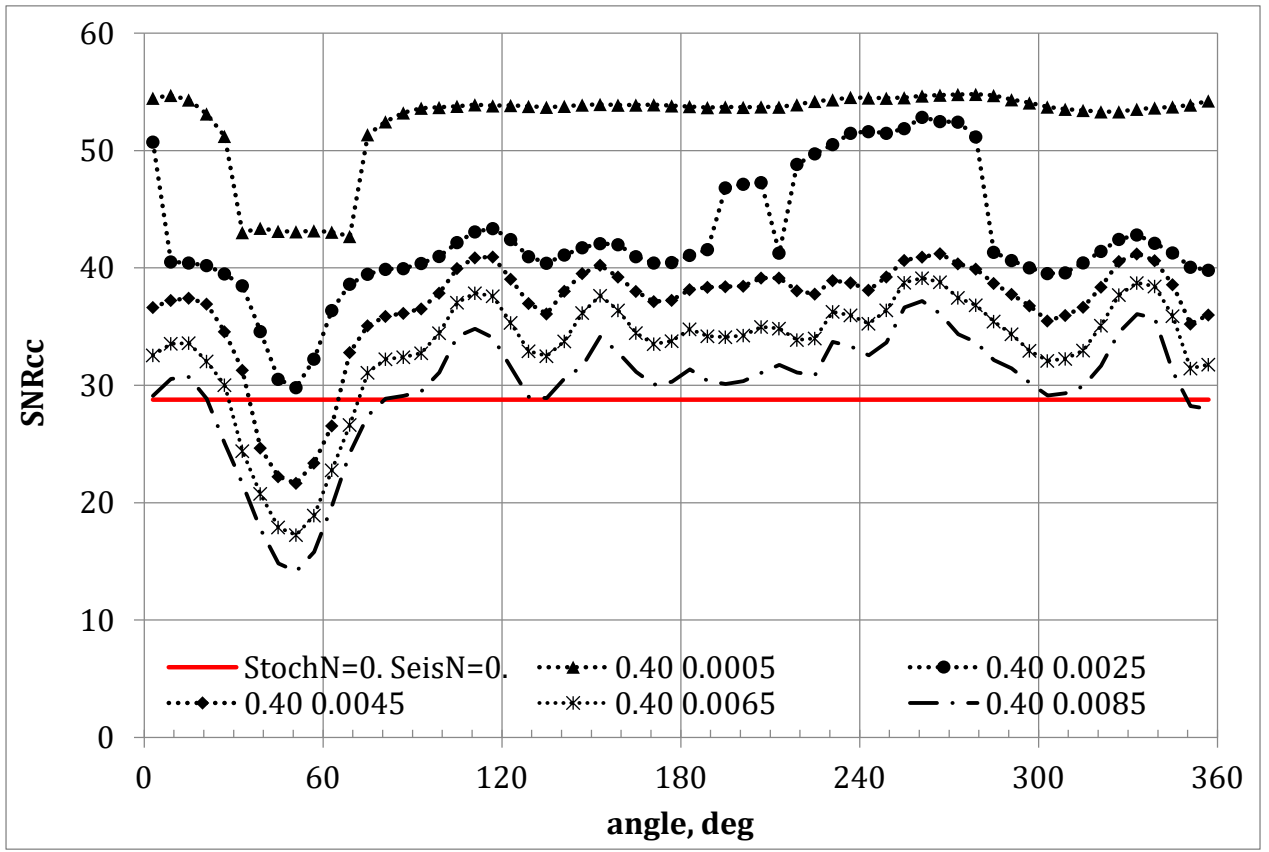

Figure 6. Same as in Figure 5 for station AKASG

As Figures 5 and 6 show, the optimal *StochN* value depends on the direction and amplitude of the transient noise component represented by the signal from the Tohoku earthquake. The ratio of an SNRcc curve for a given *SeisN*/*StochN* configuration and the "no-noise" case can be used to select the optimal configuration depending on the angle as Figure 7 shows for station AKASG and the 16.03.2022 earthquake. There are eight pairs of *SeisN*/*StochN* presented: *SeisN*=0.0005, 0.0025, 0.0045, 0.0085; *StochN* =0.1, 0.4. Since the curves with different *StochN* and the same *SeisN* values cross each other, the optimal *StochN* value depends on the angle. For the range from 70° to 360°, *StochN*=0.1 is better than 0.4, and the latter suppresses not only noise by also the signal. For angles near 55°, *StochN=0.4* gives better results and this is the trough related to the collinear sought signal and transient noise directions. The coherent ambient noise is enhanced by the transient one. To suppress the increased coherent component, a larger *StochN* value is needed.

The ratio of SNRcc values for the same *SeisN* and various *StochN* values as a function of angle between the sought signal and transient noise is shown in Figures 8 (AKASG) and 9 (CMAR). It is difficult to predict the optimal *StochN* for a given *SeisN* as it depends on the angle. The regular high-amplitude seismic phase is ambivalent in nature. It can serve for the coherent noise reduction as well as amplification.

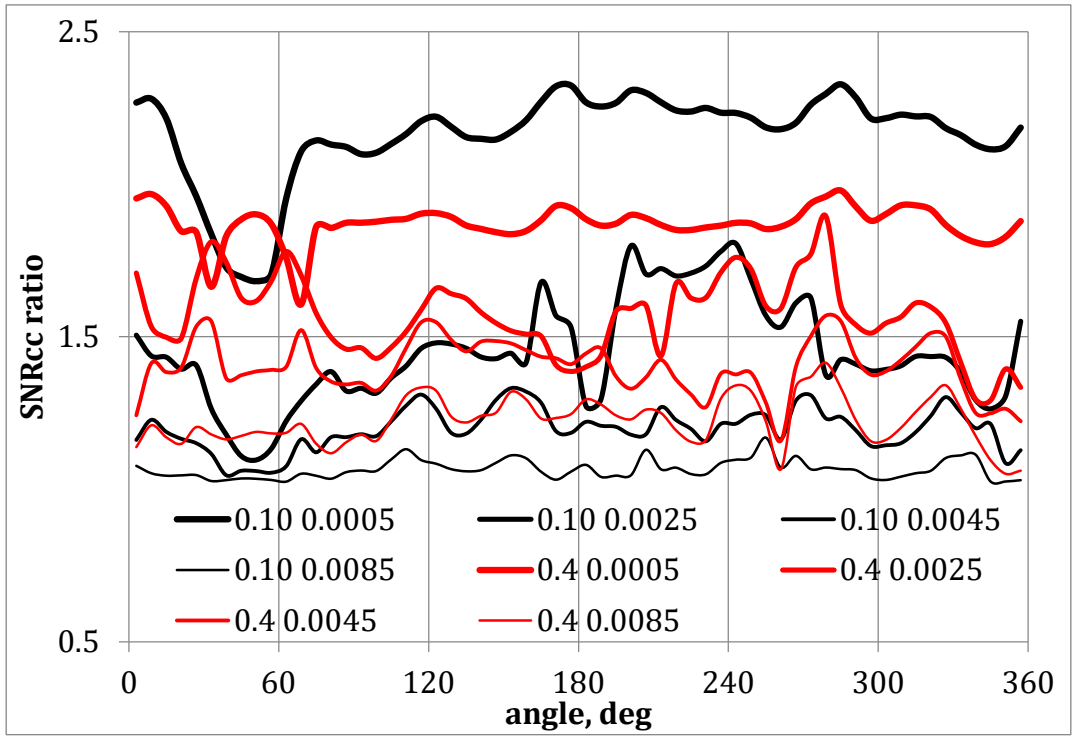

Figure 7. The SNRcc curves normalized to the "no-noise" line. Station AKASG. Eight pairs of *SeisN*/*StochN* presented: *SeisN*=0.0005, 0.0025, 0.0045, 0.0085; *StochN* =0.1, 0.4.

a)

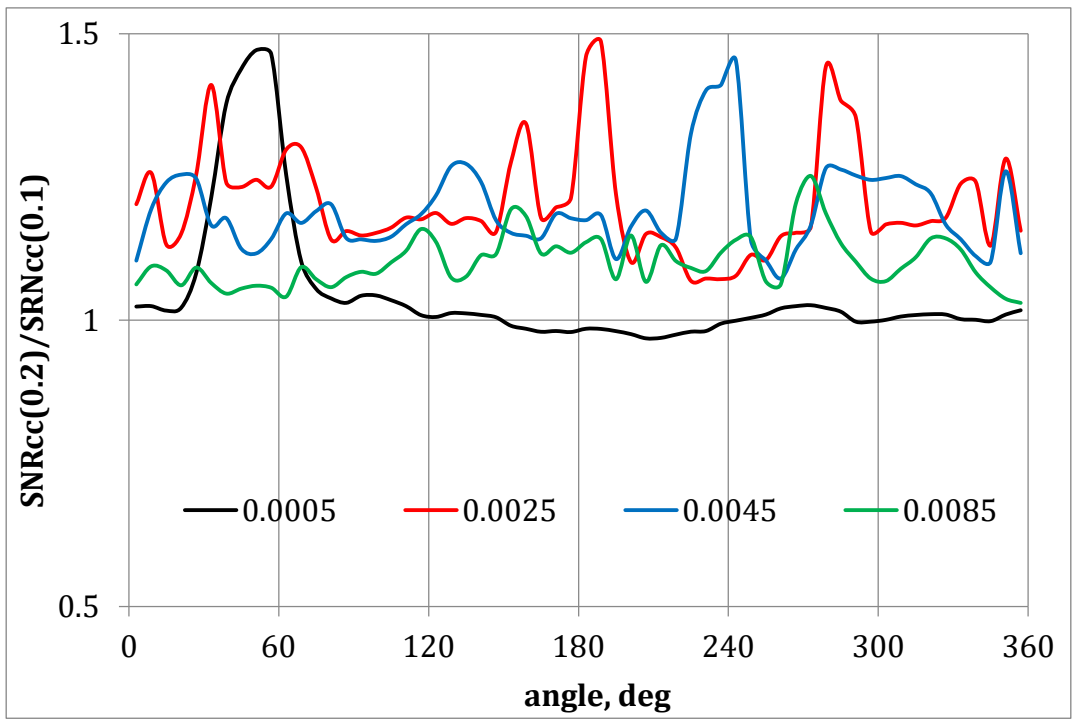

b)

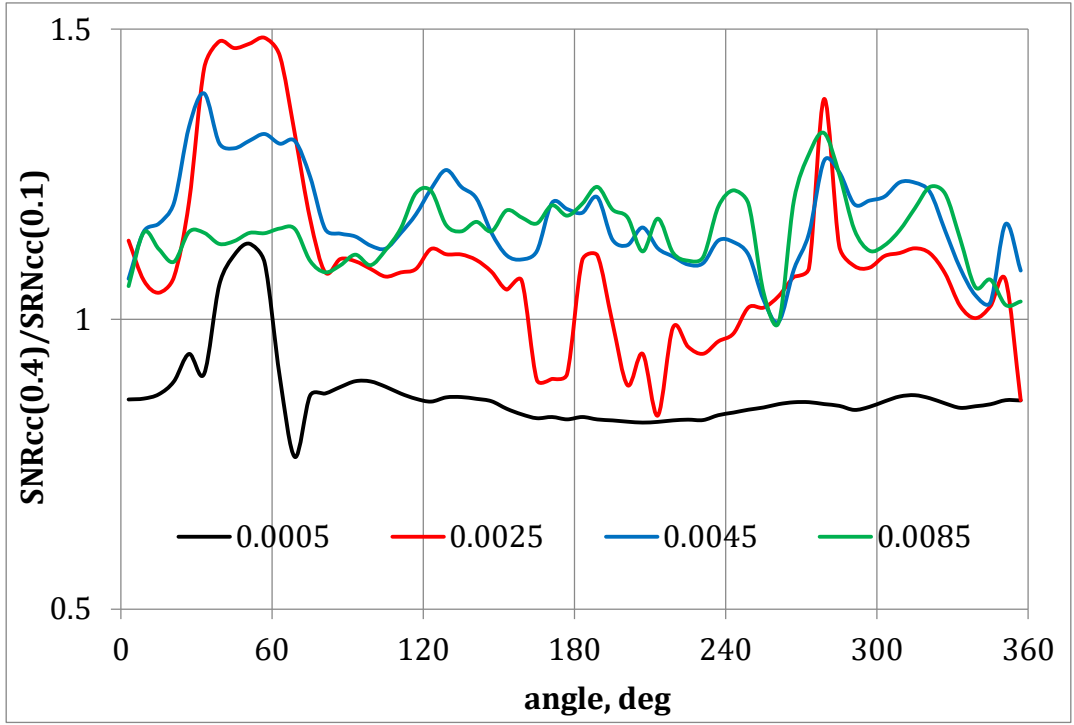

Figure 8. The ratio of two SNRcc curves a) *StochN*=0.2, *StochN*=0.1 and b) StochN=0.4, *StochN*=0.1 for various SeisN values from 0.0005 to 0.0085 as a function of Tohoku noise incidence angle. The peak ratio depends on incidence angle and *SeisN* level. The optimal *StochN* is difficult to predict for unknown noise characteristics and the whole range has to be used to find the best SNRcc and thus detect the sought signal.

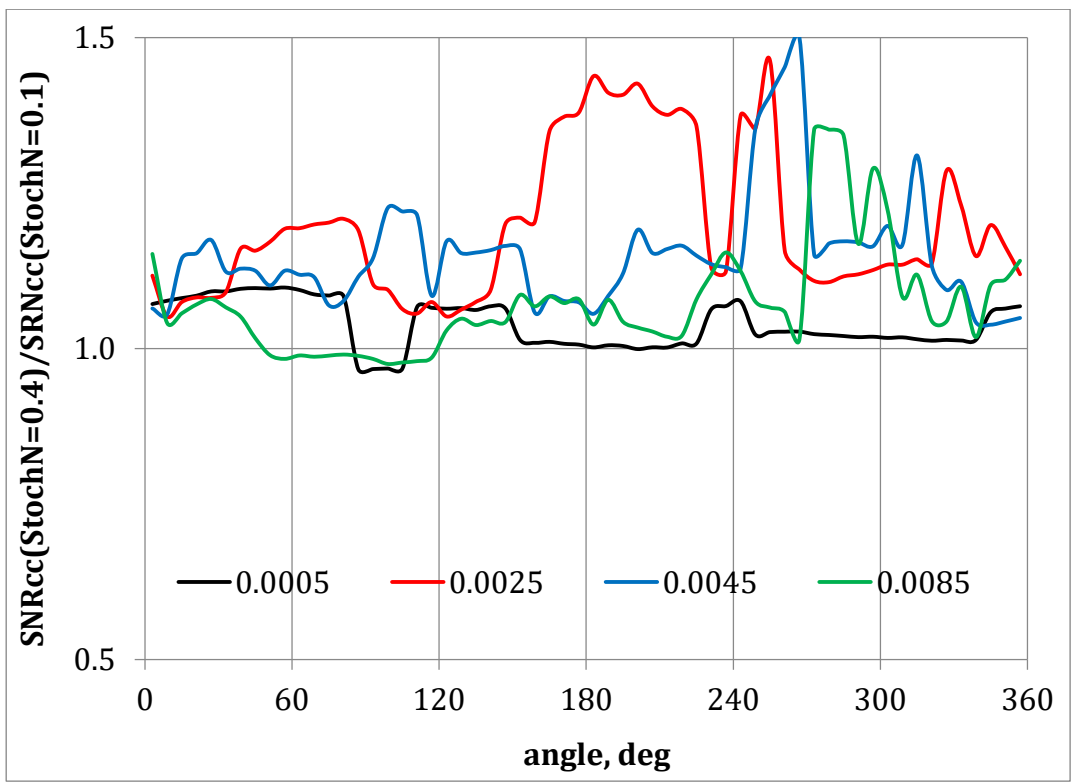

b)

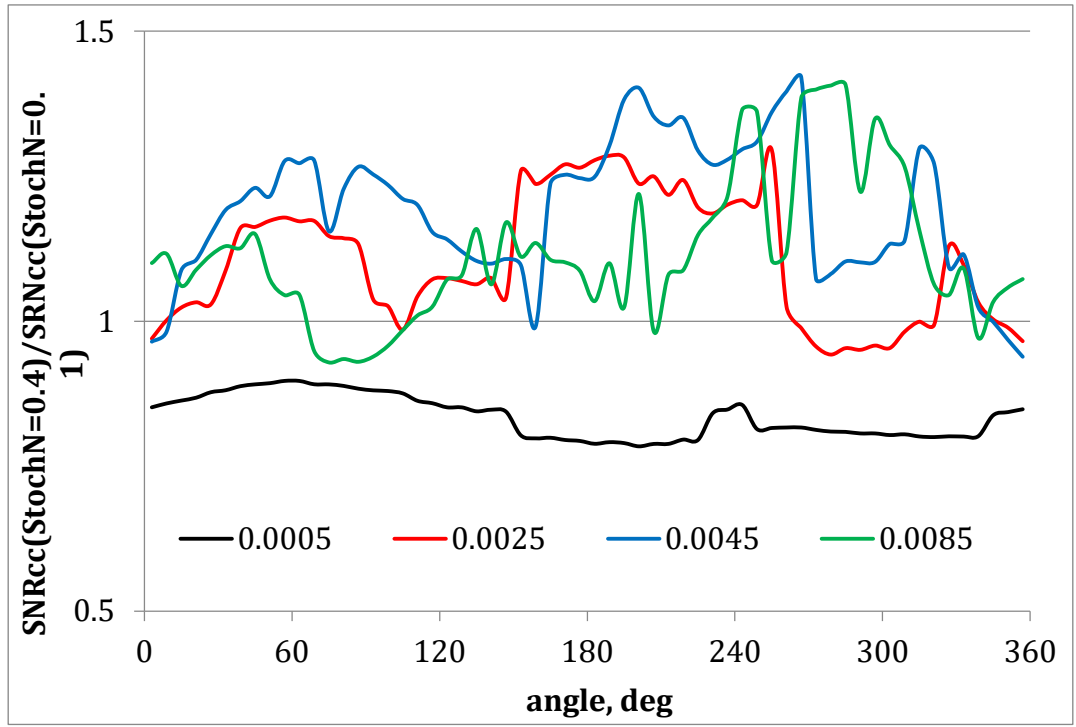

Figure 9. Same as in Figure 8 for station CMAR.

In practice, the matched filter detector has to be tuned to the parameters of actual wavefield. For simple cases of quasi-stochastic ambient seismic noise without any significant component coherent to a given template, no special noise suppression is needed. Stochastic noise added to the original waveforms can be easily tuned by varying its amplitude in the range of the potential interference of the noise. A transient noise signal with high amplitude is able to suppress the component of the ambient noise coherent to the template. However, the computer-generated stochastic noise is likely more effective in noise suppression in the cases of the interaction of the transient and ambient noise presented in Figures 7 through 9. This interaction has to be studied in order to obtain the optimal characteristics of stochastic noise when the transient noise creates negative effect on the matched filter detector. There are many cases from the noise from the almost collocated earlier earthquake on March 16, 2022 to the case of the Tohoku 2011 mega-earthquake that affected seismic detection on the global level.

*Station KSRS. March 16, 2022 earthquake as a template, sought signal, and noise signal*

Let's consider all possibilities of noise suppression/amplification at one station. The earthquake with $m_b$(IDC)=6.05 occurred on March 16, 2022 is a good example as the pre-signal noise is generated by an earlier earthquake with a lower magnitude ($m_b$(IDC)=5.22) (see Table 1). Therefore, the ambient microseismic noise near the P-wave arrival time does include a strong (likely the strongest among the others) component coherent to the sought signal. This earthquake also generated signals which have large SNR values at the IMS array stations and thus can be used as waveform templates. Therefore, the identical template and sought signal highly-coherent to the pre-signal noise create the best case scenario for the investigation of the stochastic and actual coherent noise effect on the SNRcc and SNR evolution as a function of the pre-signal noise incidence angle. As in the previous examples, one can turn the wave-plane direction of the regular phase representing the ambient noise. Regional IMS array station KSRS (37.442°N, 127.884°E, aperture ~10 km) is selected. The station-sought event distance is of 10.85° and the station-event azimuth is ~85°.

The "no-noise" case is a reference to measure the change in the SNRcc and SNR. For the P-wave detected at station KSRS, our estimate is SNR=20.6 (SNR(IDC)=32.4). The sought event was at a depth of 59 km and thus generated a direct P-wave. When this signal is used as a template for the WCC-detector, SNRcc=21.83. These values of SNR and SNRcc are both considered as constants. The SNR and SNRcc detection thresholds are 4.0 and 3.5, respectively. When the SNR and SNRcc curve fall below these thresholds, no signal is detected by the respective detection method: beamforming and WCC.

We scale the 126 s segment before the sought signal by the multiplication factor *SeisN*, move it by 80 s ahead in time, as Figure 4 shows, and add to the original recordings. The procedure is applied to each channel of KSRS separately. This makes the coherent noise to be before and within the sought signal. The SNR and SNRcc estimates for varying *SeisN* and the pivot angles between 3° and 363° are shown in Figure 10. With the increasing amplitude of the coherent noise (*SeisN* from 5 to 20), the SNRcc/SNR ratio is also increasing. This demonstrates the advantage of the WCC detection before the beamforming in the presence of coherent noise. Despite a deep trough around 85°, the SNRcc is above the "no-noise" line in a wide range of angles for the *SeisN*=5 case. Another shallow trough in the SNR is observed at 265°, i.e. arrival from the opposite to the station-event direction. Some coherency is observed even in the waveforms propagating in the opposite direction and having the opposite time delays between the channels.

The next step is to add the stochastic noise component with the amplitude scaled by *StochN*. Figure 11 is similar to Figure 5 and presents several *SeisN/StochN* pairs for station KSRS following the case in Figure 10. With increasing *SeisN,* the gain obtained from the stochastic component decreases. For *SeisN=5.0* and 10, *StochN=0.07* allows the SNRcc to be above the "no-noise" line everywhere except the range around the station-event azimuth of 85°. For *StochN=0.1,* positive gain is possible only for *SeisN=5.0*. For *StochN=0.2,* no positive gain is possible for the tested *StochN* values, except a small range around 270°.

a)

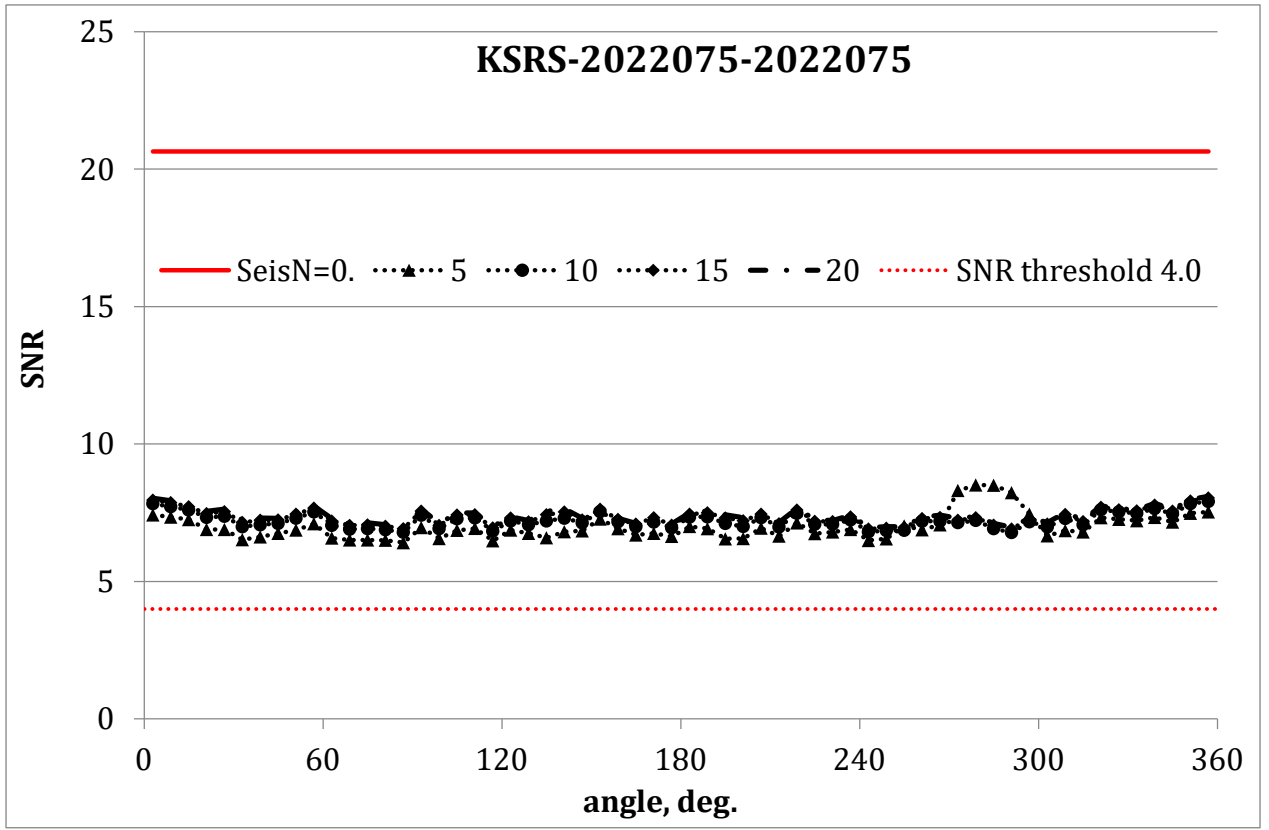

b)

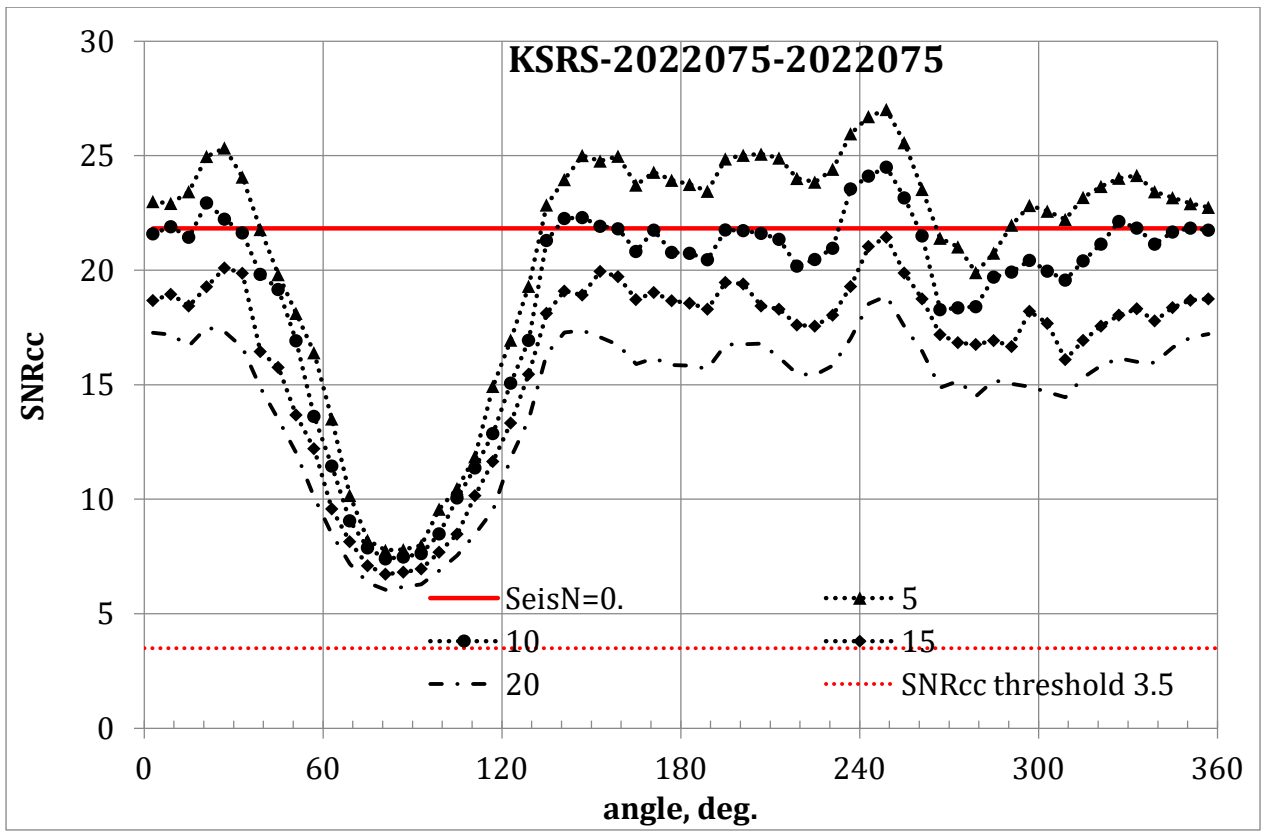

c)

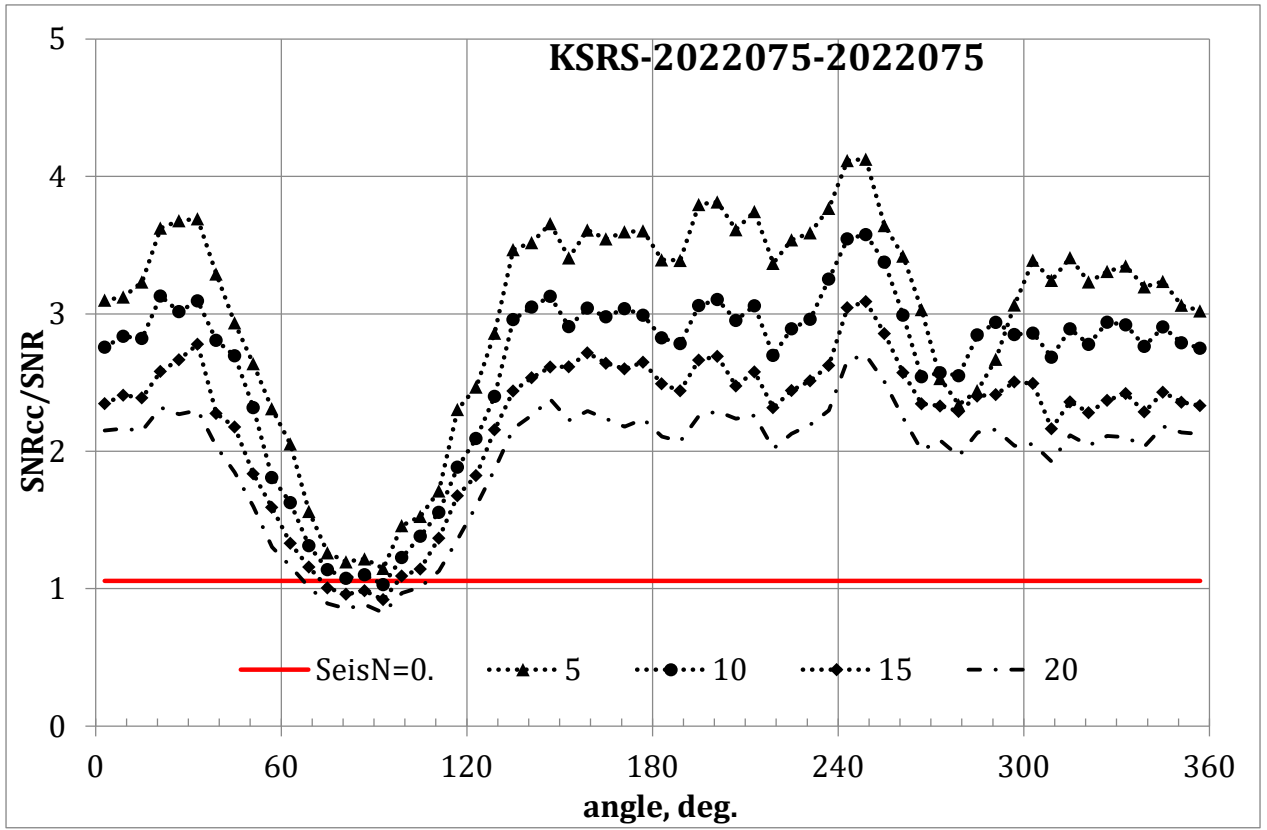

Figure 10. SNR (a) and SNRcc (b) as a function of SeisN and pivot angle. c) The SNRcc/SNR ratio.

Figure 11 illustrates the advantages and limitation of the transient and stochastic noise suppression. The high-amplitude transient noise coherent to the template is a very effective measure to hide the signals of interest. For the actual amplitude ratio of ~10 of the sought and noise signals at KSRS, the *SeisN*=10 case (equal signal and noise amplitudes) is still possible to resolve with the *StochN*=0.07. For larger *SeisN*, the sought signal disappears in the coherent noise. If to swap the events, with the larger occurring 126 s before the weaker, the latter would not be detected at KSRS. Such an effect is likely working for the immediate aftershocks of the major earthquakes as Figure 12 demonstrates. There is a also still period of 10 to 20 minutes after such earthquake without aftershocks. The WCC method is able to reduce the detection threshold during these first minutes after the major earthquake as shown in [Kitov *et al*., 2026b]. This is an important extension of the beamforming method as the largest aftershocks of the mega-earthquakes are usually missing from the global catalogs. There are many aftershocks in the first hour after the Tohoku, Sumatra, and Kamchatka earthquakes to be added to these catalogs.

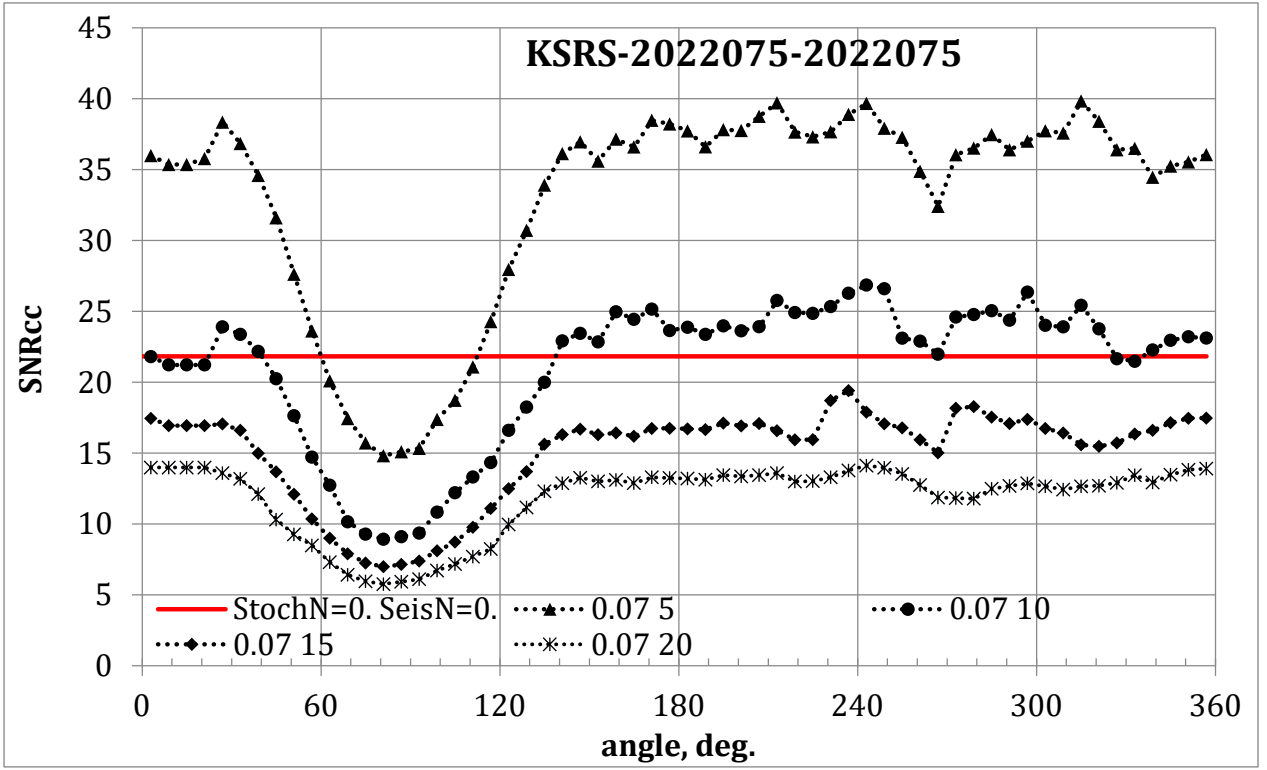
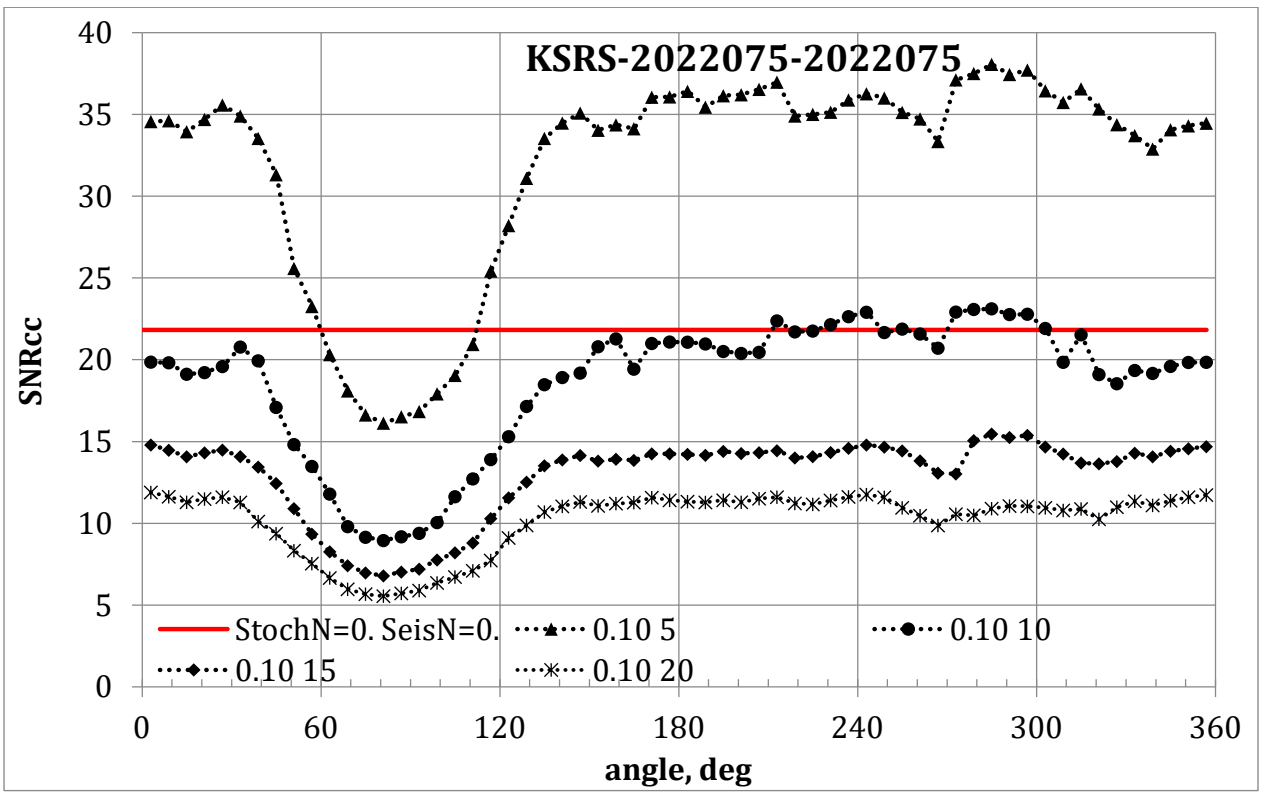

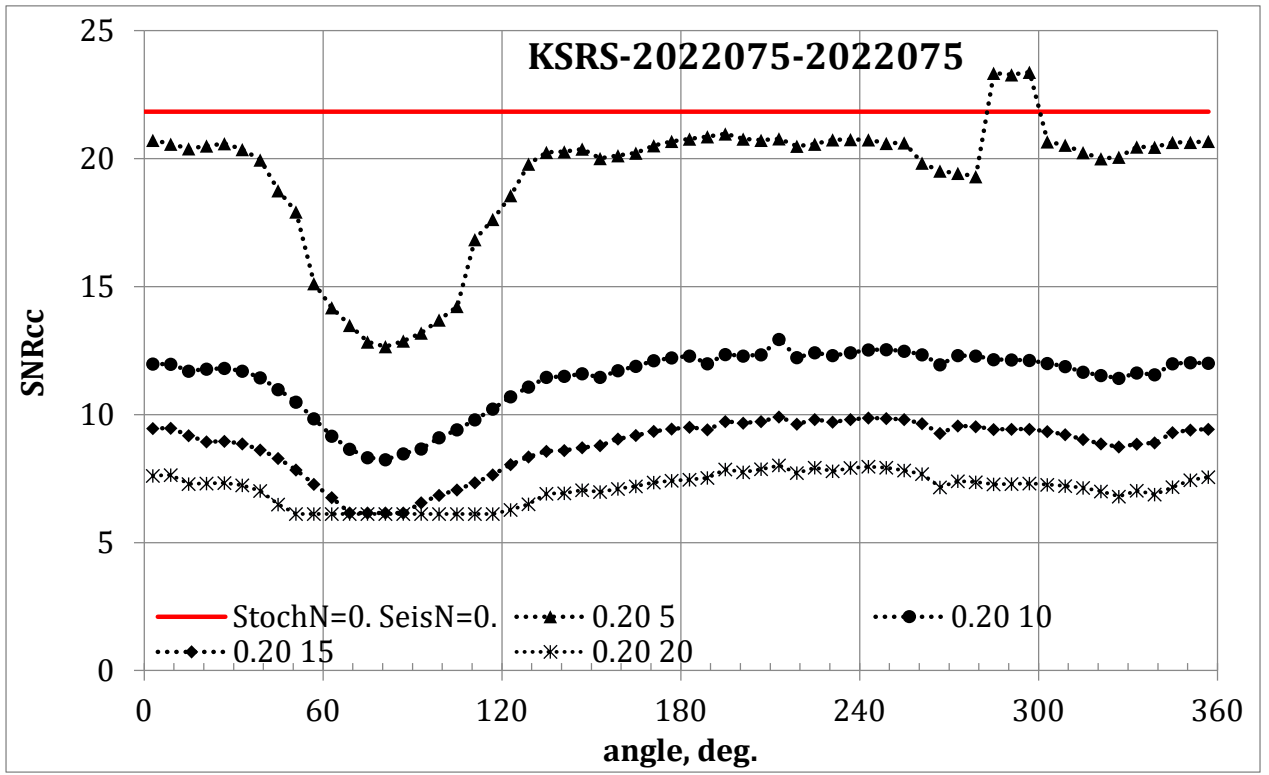

Figure 11. SNRcc dependence on the plane (noise) wave angle for various *SeisN*/*StochN* pairs

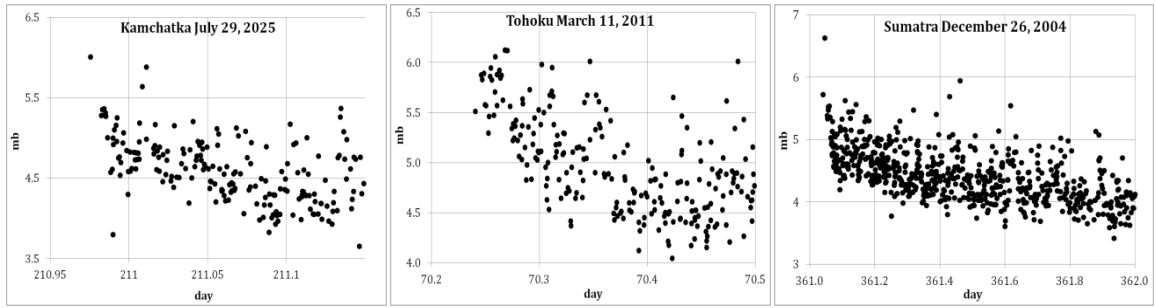

Figure 13. The evolution of aftershock sequence for major earthquakes: Kamchatka July 25, 2025, Tohoku March 11, 2011, Sumatra December 26, 2004. Origin times and magnitudes are borrowed from the REB.

Figure 14 presents the SNRcc/SNR ratios for two limit cases in Figure 11: *StochN*=0.07 and 0.2. This ratio characterizes the gain obtained by the WCC method relative to beamforming for the same *SeisN* values, i.e. in the case of actual noise to be the same as the added transient signal. Such an estimate is important for the cases when a Tohoku-like earthquake increases the noise level by orders of magnitude worldwide. The transient signal is considered here as the noise hiding the signals of interest rather than the noise suppressor.

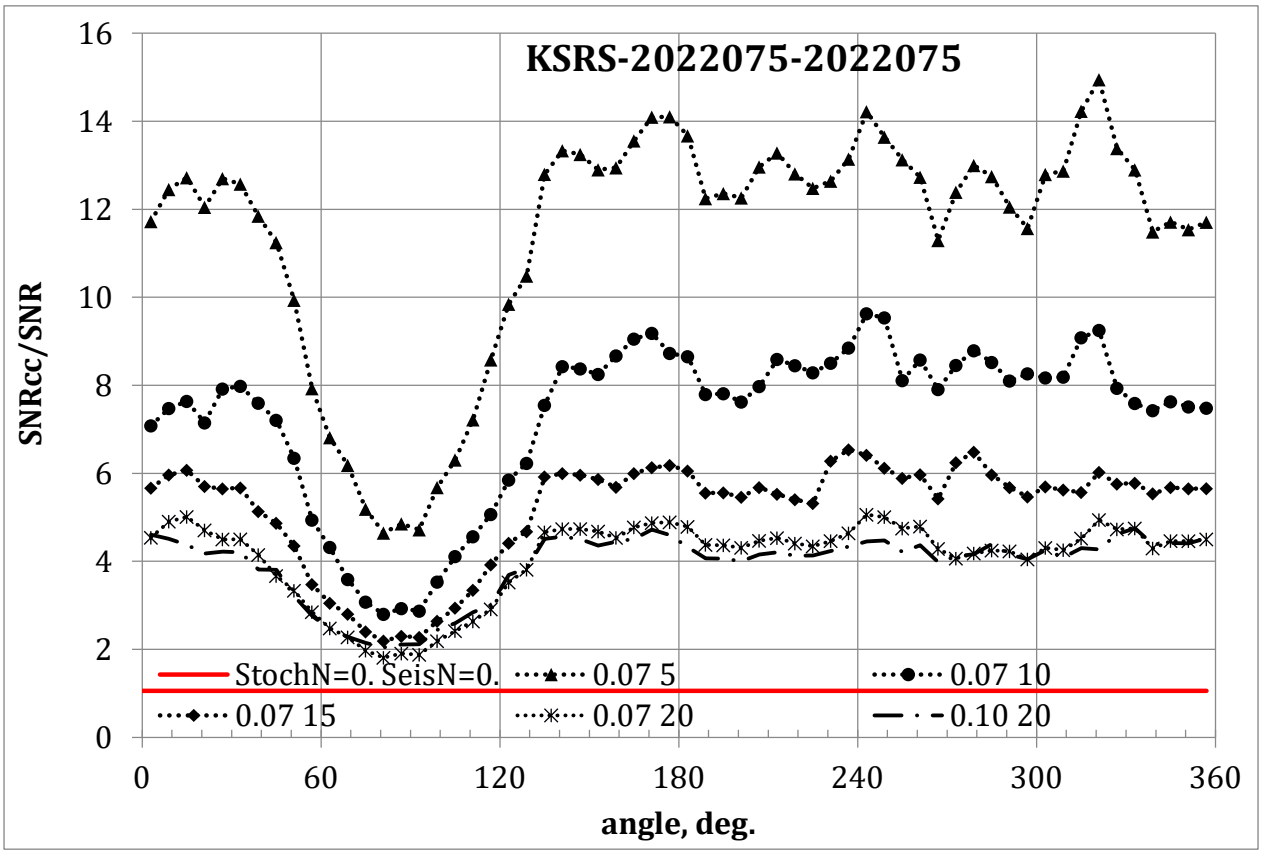
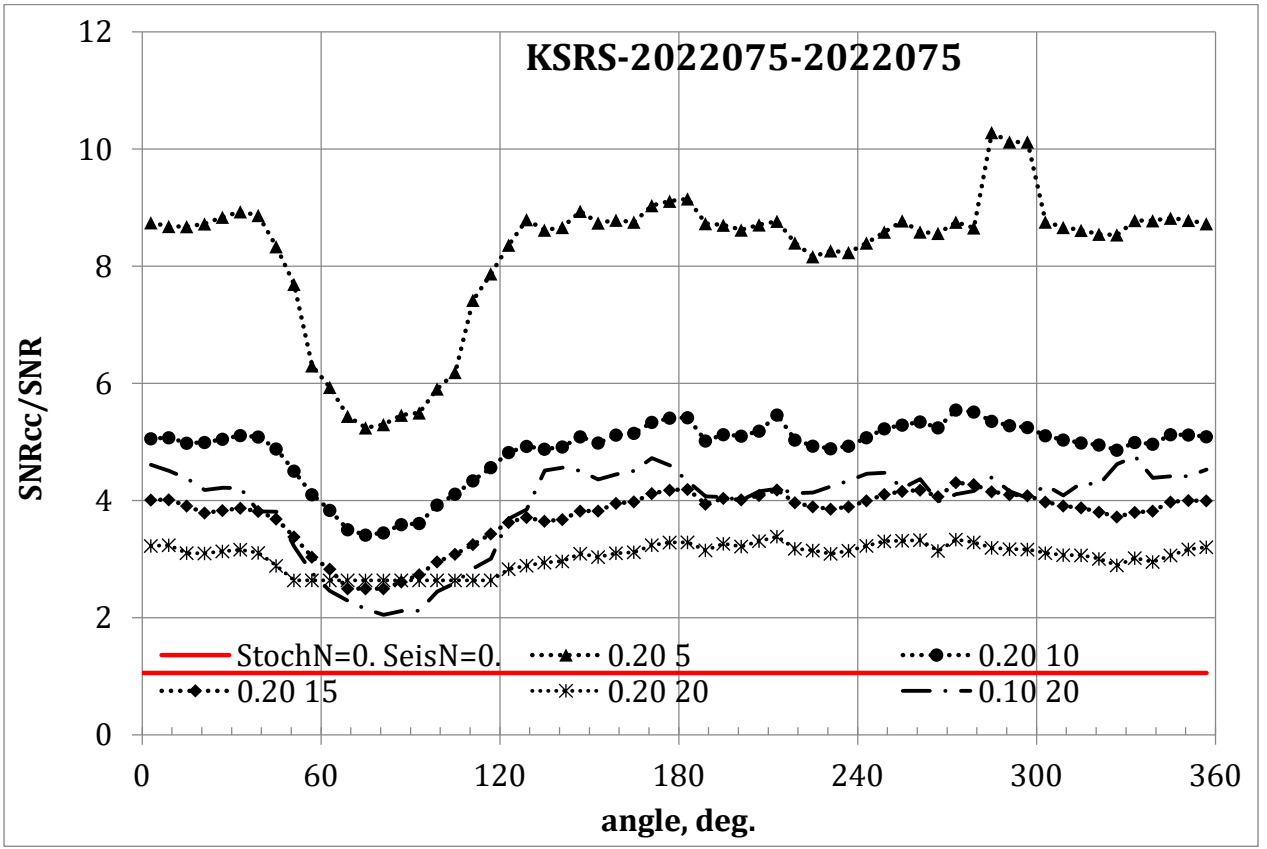

Figure 14. Ratios SNRcc/SNR for two limit cases in Figure 11.

Overall, the WCC method has an advantage of a factor of 10 to 15 for *SeisN*=5.0. The gain is positive, i.e. above the "no-noise" line, in all cases in Figure 14. The angle dependence is higher for lower *SeisN* values and practically disappears for *StochN*=0.2 and *SeisN*=20, where the gain is of around 3.

Figure 15 shows SNRcc ratio for various *SeisN*/*StochN* cases and *StochN* =0.0 case. This presents the input of stochastic noise to the gain in SNRcc. For high-amplitude seismic noise, there is no gain. The results are very similar to those presented in Figure 7. The peak ratio is near the station-event azimuth and the opposite direction. For the *StochN*=0.10 and *SeisN*=5 case, the peak value is above 2.0 and around 1.5 for the other angles. The gain is decreasing with *SeisN* increase and falls below 1.0 for *SeisN* above 15 in line with Figure 11.

Figure 16 shows the ratio of SNRcc values for *StochN* of 0.07 and 0.1 as a function of *SeisN* and angle. For *SeisN*=5.0, the ratio is above 1.0 for *StochN*=0.07 for almost all angles except two narrow intervals around 85° and 265°. The *StochN*=0.10 case is optimal for all other *SeisN* values and angles as the ratio is below 1.0. Therefore, the optimal *StochN* value depends on the other parameters for this specific configuration of sought signal, template and transient noise. For all other configurations, a grid search has to be applied to find the optimal values.

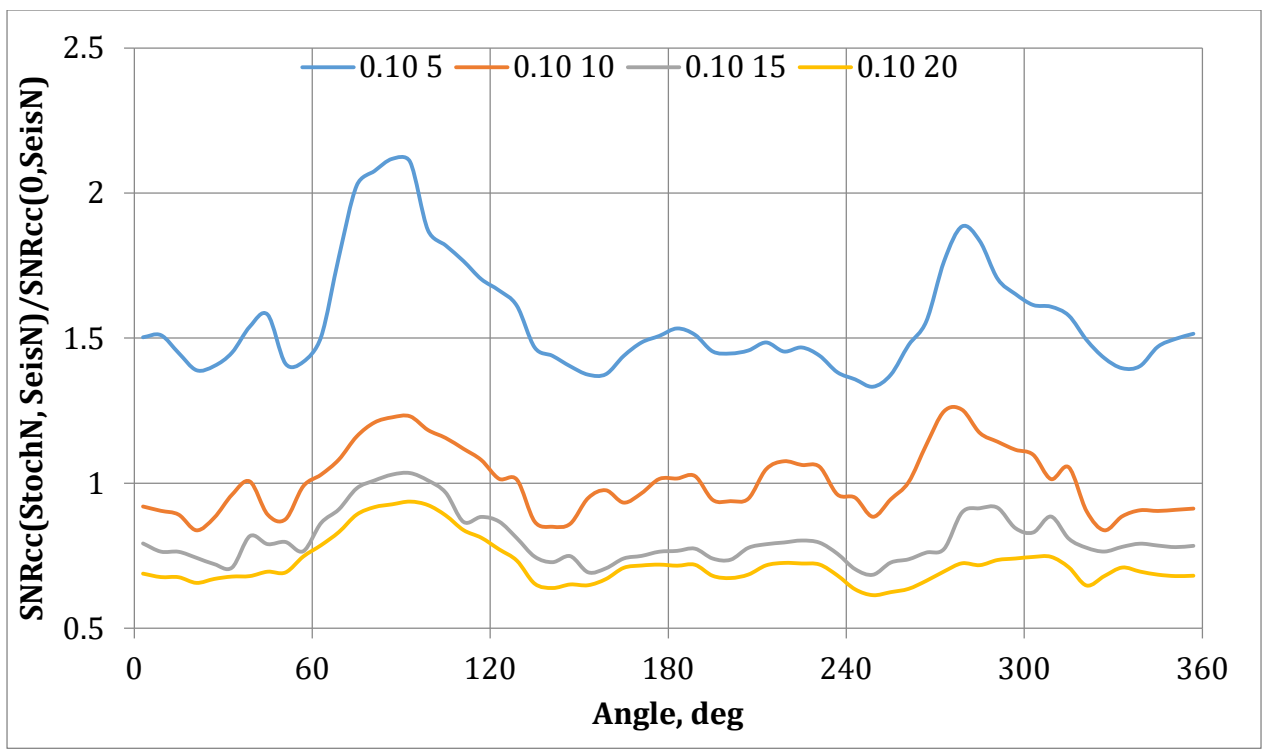

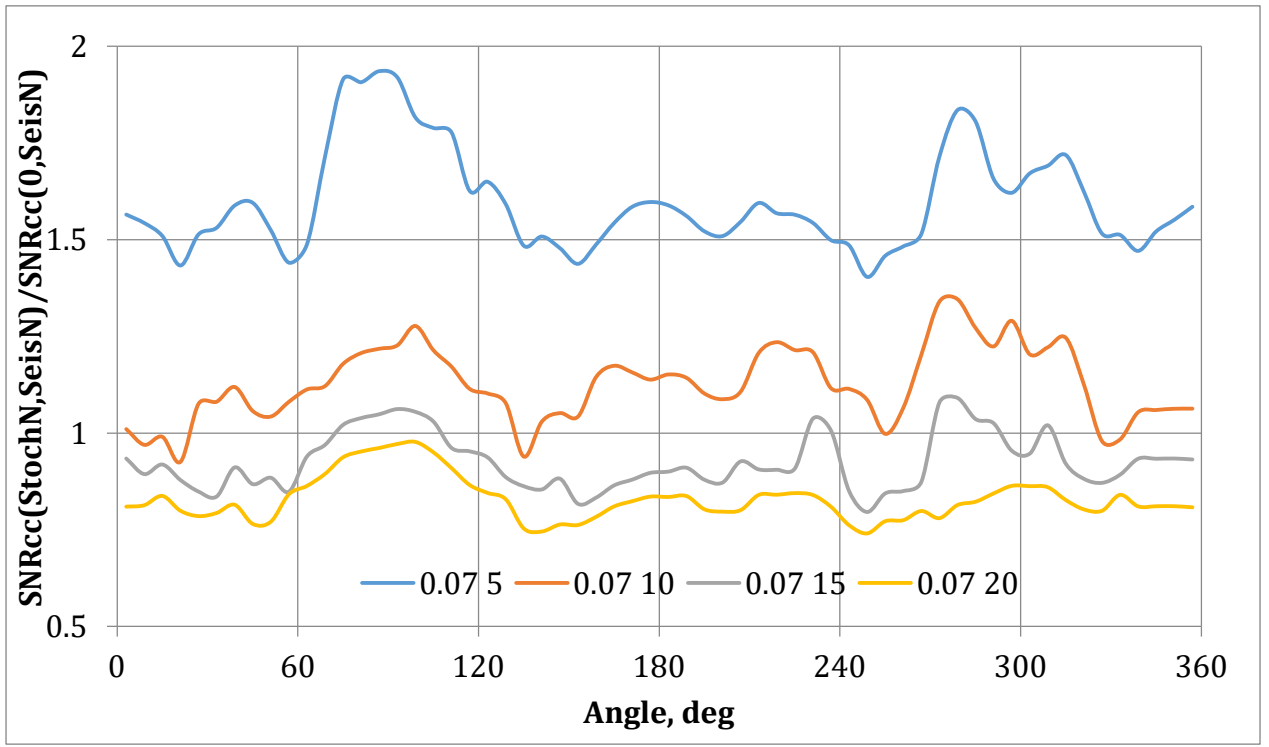

Figure 15. The SNRcc ratio for *SeisN-StochN* pairs to the case with *StochN*=0.

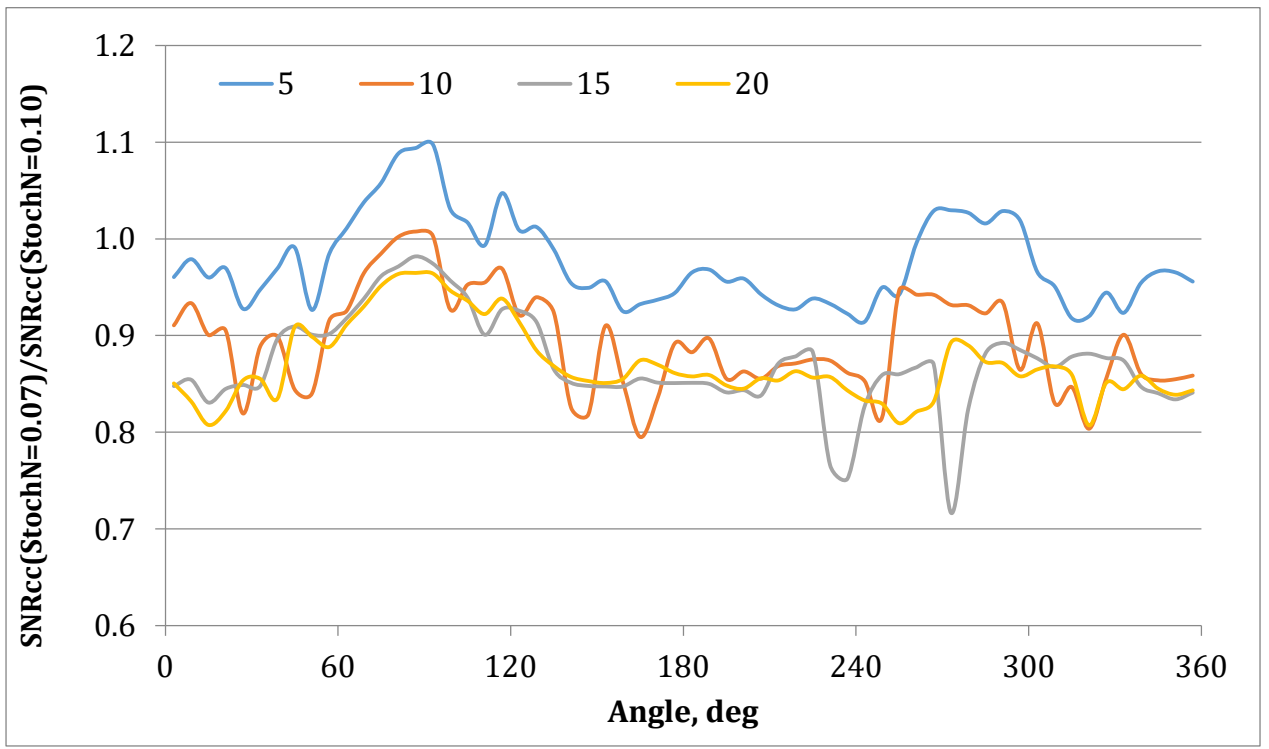

Figure 16. SNRcc ratios for *StochN*=0.07 and *StochN*=0.1 for various *SeisN* and angles.

**Discussion**

We have estimated the performance of the beamforming and the WCC methods in two cases related to the sources near the Tohoku 2011 earthquake. The WCC method demonstrates significant advantage in detection of signals in the coherent noise environment. This is the most difficult case for the beamforming as its major assumption is the stochastic noise behavior allowing destructive noise interference without any visible loss in the sought signal amplitude. The WCC method is able to reduce the detection threshold but also suffers the coherency problem. There are two ways to suppress the coherent noise component to enhance the WCC detection method - to add a transient signal not coherent to the sough signals or/and add stochastic noise component to the actual data. The choice of the optimal method depends on the case. Therefore, the case by case consideration is the only possible approach to obtain the optimal detection conditions. When obtained, the optimal processing parameters can decrease the detection threshold by order of magnitude even in the worst case scenario of the full coherence of the noise and the sought signal.

There are many features related to the stochastic noise and regular signals used as noise to be included in the in-depth analysis tool. These features have to provide flexibility in the selection of the optimal parameters. Individual and joint effects of all parameters have to be studied in advance and tuned to the near-optimal values to simplify the user control. There are any problems and questions to answer. *Inter alia*:

- How does adding real transient noise affect SNRcc and SNR depending on the angle? The noise can be coherent or incoherent to the sought signal.
- How does adding real transient noise increase the SNRcc/SNR ratio?
- How does noise affect the detection threshold for SNRcc and SNR, if it arrives during teh sought signal arrival to the station. For example, how the detection of a signal from an underground explosion will be affected by a signal from a Toholu size earthquake?
- What is the best procedure to add stochastic noise? How does the procedure of stochastic noise generation affect its performance?
- How does stochastic noise affect SNRcc and SNR and the SNRcc/SNR ratio? In this case, detection thresholds are more important than increasing the ratio.

**References**

*Adushkin V.V., Kitov I.O., Konstantinovskaya N.L., Nepeina K.S., Nesterkina M.A., and Sanina I.A*. Detection of ultraweak signals on the Mikhnevo small-aperture seismic array by using cross-correlation of waveforms // Dokl. Earth Sci. 2015. V. 460. № 2. P. 189–191.

*Adushkin V.V., Bobrov D.I., Kitov I.O., Rozhkov M.V., Sanina I.A.* Remote detection of aftershock activity as a new method of seismic monitoring // Dokl. Earth Sci. 2017. V. 473. № 1. P. 303–307.

*Adushkin V. V., Kitov I. O., and Sanina I. A.* Further Development of the Matched Filter Method for Solving Seismological Problems // Doklady Earth Sciences. 2025a. V. 523:13. https://doi.org/10.1134/S1028334X25606182

*Adushkin V. V., Kitov I. O., and Sanina I. A.* Detection of weak aftershocks at regional and teleseismic distances // J. Volcanology and Seismology. 2025b. (in press)

*Arrowsmith S. J., Eisner L.* A technique for identifying microseismic multiplets and application to the Valhall field, North Sea // Geophysics. 2006. V. 71. P. 31–40.

*Baisch S., Ceranna L., Harjes H.-P.* Earthquake Cluster: What Can We Learn from Waveform Similarity? // Bull. Seismol. Soc. Am. 2008. V. 98. P. 2806–2814.

*Beaucé, E., W. B. Frank, L. Seydoux, P. Poli, N. Groebner, R. D. van der Hilst, and M. Campillo* (2023). BPMF: A Backprojection and Matched-Filtering Workflow for Automated Earthquake Detection and Location, Seismol. Res. Lett. 95, 1030–1042, doi: 10.1785/0220230230./

*Bobrov D., Kitov I., Zerbo L.* Perspectives of cross correlation in seismic monitoring at the International Data Centre // Pure Appl. Geophys. 2014. V. 171. № 3–5. P. 439–468.

*Bobrov D.I., Kitov I.O., Rozhkov M.V., Friberg P.* Towards Global Seismic Monitoring of Underground Nuclear Explosions Using Waveform Cross Correlation. Part I: Grand Master Events // Seismic Instruments. 2016a. V. 52. № 1. P. 43–59.

*Bobrov D.I., Kitov I.O., Rozhkov M.V., Friberg P.* Towards Global Seismic Monitoring of Underground Nuclear Explosions Using Waveform Cross Correlation. Part II: Synthetic Master Events // Seismic Instruments. 2016b. V. 52. № 3. P. 207–223.

*Bobrov D., Kitov I., Rozhkov M.* Studying seismicity of the Atlantic Ocean using waveform cross-correlation // NNC RK Bulletin. 2017. 2(70). P. 5-19.

Comprehensive Nuclear-Test-Ban Treaty. 1996. Protocol to the Comprehensive Nuclear-Test-Ban Treaty. https:www.ctbto.org.sites.default.files.Documents.treatytext.tt.html

*Coyne J., Bobrov D., Bormann P., Duran E., Grenard P., Haralabus G., Kitov I., Starovoit Yu.* Chapter 15: CTBTO: Goals, Networks, Data Analysis and Data Availability / Ed. P. Bormann // New Manual of Seismological Practice Observatory. 2012. P. 1–41. DOI: 10.2312/GFZ.NMSOP-2_ch15

*Gibbons S., Ringdal F.* A waveform correlation procedure for detecting decoupled chemical explosions // NORSAR Scientific Report: Semiannual Technical Summary № 2. NORSAR, Kjeller, Norway. 2004. P. 41–50.

*Gibbons S.J., Ringdal F.* The detection of low magnitude seismic events using array based waveform correlation // Geophys. J. Int. 2006. V. 165. P. 149–166.

*Gibbons S.J., Sorensen M.B., Harris D.B. and Ringdal F.* The detection and location of low magnitude earthquakes in northern Norway using multichannel waveform correlation at regional distances // Phys. Earth Planet.Inter. 2007. Vol. 160. N. 3–4. P. 285–309.

*Gibbons S.J., Schweitzer J., Ringdal F., Kværna T., Mykkeltveit S. and Paulsen B.* Improvements to seismic monitoring of the European Arctic using three component array processing at SPITS // Bull. Seismol. Soc. Am. 2011. Vol. 101. N. 6. P. 2737–2754.

*Gibbons S.J., Pabian F., Näsholm S.P., Kværna T., Mykkeltveit S.* Accurate relative location estimates for the North Korean nuclear tests using empirical slowness corrections // Geophys. J. Int. 2017. V. 208. № 1. P. 101–117.

*Herrmann, M., T. Kraft, T. Tormann, L. Scarabello, Wiemer S.* A consistent high-resolution catalog of induced seismicity in Basel based on matched filter detection and tailored post-processing // J. Geophys. Res. 2019. V. 124. N 8. P. 8449–8477, doi: 10.1029/2019JB017468.

*Israelsson H.* Correlation of waveforms from closely spaced regional events // Bull. Seismol. Soc. Am. 1990. V. 80. № 6. P. 2177–2193.

*Joswig M.* Pattern recognition for earthquake detection // Bull. Seismol. Soc. Am. 1990. V. 80. P. 170–186.

*Joswig M., Schulte-Theis H.* Master-event correlations of weak local earthquakes by dynamic waveform matching // Geophys. J. Int. 1993. V. 113. Issue 3. P. 562–574. DOI: https://doi.org/10.1111/j.1365-246X.1993.tb04652.x

*Kim W.-Y., Richards P., Schaff D. et al.* Identification of Seismic Events on and near the North Korean Test Site after the Underground Test Explosion of 3 September 2017 // Seismol. Res. Lett. 2018. V. 89. № 6. P. 2120–2130.

*Kitov I.O., Sanina I.A.* Analysis of Sequences of Aftershocks Initiated by Underground Nuclear Tests Conducted in North Korea on September 9, 2016 and September 3, 2017 // Seism. Instr. 2022. V. 58. P. 567–580.

*Kitov I.O., Rozhkov M.V.* New Applications at the International Data Centre for Seismic, Hydroacoustic and Infrasound Expert Technical Analysis. Vienna. Austria / Eds M. Kalinowski et al. // Twenty-five Years Progress of the Comprehensive Nuclear-Test-Ban Treaty Verification System. PTS Preparatory Commission for the CTBTO. 2024. P. 233–254.

*Kitov I.O., Dricker I.* Enhancing Signal Detection in High-Amplitude Seismic Noise Using Arrays, Waveform Cross-Correlation, and Noise Whitening // AGU 2025. New Orleans; December 15, 2025; Poster S13E-VR8868.

*Kitov I.O., Sanina I.A.* Recovery of the Aftershock Sequence of the North Atlantic Earthquake Using Waveform Cross-Correlation // Geodynamics & Tectonophysics. 2025a. 16 (4), 0838. doi:10.5800/GT-2025-16-4-0838

*Kitov I.O., Sanina I.A.* Application of regular phases to suppress coherent noise in waveform cross-correlation on a seismic array // Seismic Instruments. 2025b. doi.org/10.21455/si2025.4-2

*Kitov I. O., Sanina I. A., Volosov S. G., Konstantinovskaya N. L.* The 20th Anniversary of the Installation of the Small-Aperture Mikhnevo Array for Monitoring Induced Seismicity // 2025. Izvestiya, Physics of the Solid Earth. V. 61. N. 2. P. 288–304. DOI: 10.1134/S1069351325700181

*Kitov I.O., Sanina I.A., Sokolova I.N.* Estimation of the detection threshold of seismic events in the noise of the Tohoku earthquake // J. Volcanology and Seismology. 2026a. (in press)

*Kitov I.O., Sanina I.A., Sokolova I.N, Vinogradov Yu.A.* Study of Seismic Activity between the Mainshock and the First Aftershock of the Kamchatka Earthquake on July 29 2025 // Russian Journal Earth Sciences. 2026b. (in press)

*Mesimeri, M., D. Armbruster, P. Kästli, L. Scarabello, T. Diehl, J. Clinton, and S. Wiemer* (2024). SCDetect: A SeisComP Module for Real-Time Waveform Cross-Correlation-Based Earthquake Detection, Seismol. Res. Lett. 95, 1961–1975, doi: 10.1785/0220230164.

*Mu, D., E.-J. Lee, and P. Chen* Rapid earthquake detection through GPU-Based template matching. // Comput. Geosci. 2019. V. 109, P. 305–314. doi: 10.1016/j.cageo.2017.09.009.

*Ringdal F.* GSETT 3: a test of an experimental international seismic monitoring system //Annals of geophysics. 1994. 37(3). DOI:10.4401/ag-4203

*Saragiotis C., Kitov I.* Tuning IMS station processing parameters and detection thresholds to increase detection precision and decrease detection miss rate // EGU General Assembly 2020, Online, 4–8 May 2020, EGU2020-8949. 2020. https://doi.org/10.5194/egusphere-egu2020-8949

*Schaff, D. P.* (2009). Broad-scale applicability of correlation detectors to China seismicity, Geophys. Res. Lett. 36, no. 11, 1–5, doi:10.1029/2009GL038179.

*Schaff D.P. and Richards P.G.* Repeating seismic events in China // Science. 2004. Vol. 303. P. 1176–1178.

*Schaff D. P. and Richards P.G.* Improvements in magnitude precision, using the statistics of relative amplitudes measured by cross correlation // Geophys. J. Int. Seismology. 2014. 197(1). P. 335–350. DOI: 10.1093/gji/ggt433

*Schaff, D. P., and F. Waldhauser* (2010). One magnitude unit reduction in detection threshold by cross correlation applied to Parkfield (California) and China seismicity, Bull. Seismol. Soc. Am. 100, no. 6, 3224–3238, doi: 10.1785/0120100042.

*Schaff, D.P., Kim, W.-Y. and Richards P.G.* Background Seismicity for parts of the northern Korean peninsula // 2025. CTBT Science & Technology Conference. P1.2-714. Vienna, Austria.

*Schweitzer J., Fyen J., Mykkeltveit S., Gibbons S.J., Pirli M., Kühn D. and Kværna T.* // Seismic arrays, in New Manual of Seismological Practice Observatory. Bormann. P., Ed. 2012. Ch. 9. doi: 10.2312/GFZ.NMSOP2_ch9.

*Selby, N.* Relative location of the October 2006 and May 2009 DPRK announced nuclear tests using International Monitoring System Seismometer arrays // Bull. Seismol. Soc. Am. 2010. V. 100. P. 1779-1784.

*Turin, G. L.* An introduction to matched filters // IRE Transactions on Information Theory. 1960.V. 6. PP. 311–329. doi:10.1109/TIT.1960.1057571. S2CID 5128742

vDEC. 2026. /https://www.ctbto.org/resources/for-researchers-experts/vdec

*Waldhauser, F., Schaff, D.* Large-scale cross correlation based relocation of two decades of northern California seismicity // J. Geophys. Res. 2008. V. 113. B08311. doi:10.1029/2007JB005479.